\documentclass[traditabstract]{aa}  

\usepackage{graphicx}
\usepackage{rotating}
\usepackage{natbib}
\usepackage{txfonts}
\usepackage{threeparttable}
\usepackage{float}
\usepackage{longtable}

\newcommand\Tstrut{\rule{0pt}{2.5ex}}         
\newcommand\Bstrut{\rule[-0.9ex]{0pt}{0pt}}   

\begin{document} 

\title{The CARMENES search for exoplanets around M dwarfs}

\subtitle{The enigmatic planetary system GJ 4276: \\ 
One eccentric planet or two planets in a 2:1 resonance?\thanks{Radial velocity data 
(Tables C1 and C2) are only available 
in electronic form at the CDS via anonymous ftp to \protect\url{cdsarc.u-strasbg.fr} 
(\protect\url{130.79.128.5})
or via \mbox{\protect\url{http://cdsweb.u-strasbg.fr/cgi-bin/qcat?J/A+A/}}}}

\author{
E. Nagel\inst{1}
\and S. Czesla\inst{1}
\and J. H. M. M. Schmitt\inst{1}
\and S. Dreizler\inst{2}
\and G. Anglada-Escud\'{e}\inst{3,4}
\and E. Rodr\'{i}guez\inst{3}
\and I. Ribas\inst{5,6}
\and A.~Reiners\inst{2}
\and A. Quirrenbach\inst{7}
\and P. J. Amado\inst{3}
\and J.~A.~Caballero\inst{8}
\and J. Aceituno\inst{3,9}
\and V. J. S. B\'{e}jar\inst{10,11}
\and M. Cort\'{e}s-Contreras\inst{8}
\and L.~Gonz\'{a}lez-Cuesta\inst{10,11}
\and E.~W. Guenther\inst{12}
\and T.~Henning\inst{13}
\and S. V. Jeffers\inst{2}
\and A. Kaminski\inst{7}
\and M. Kürster\inst{13}
\and M. Lafarga\inst{5,6}
\and M. J. L\'{o}pez-Gonz\'{a}lez\inst{3}
\and D.~Montes\inst{14}
\and J. C. Morales\inst{5,6}
\and V.~M.~Passegger\inst{1}
\and C. Rodr\'{i}guez-L\'{o}pez\inst{3}
\and A. Schweitzer\inst{1}
\and M. Zechmeister\inst{2}
}
   
\institute{
Hamburger Sternwarte, Gojenbergsweg 112, 21029 Hamburg, Germany\\
\email{evangelos.nagel@hs.uni-hamburg.de}
\and Universität Göttingen, Institut für Astrophysik, Friedrich-Hund-Platz 1, 37077 Göttingen, Germany
\and Instituto de Astrof\'{i}sica de Andaluc\'{i}a (IAA-CSIC), Glorieta de la Astronom\'{i}a s/n, 18008 Granada, Spain
\and School of Physics and Astronomy, Queen Mary, University of London, 327 Mile End Road, London, E1 4NS, UK
\and Institut de Ci\`{e}ncies de l’Espai (ICE, CSIC), Campus UAB, C/ de Can Magrans s/n, 08193 Cerdanyola del Vall\`{e}s, Spain
\and Institut d’Estudis Espacials de Catalunya (IEEC), C/ Gran Capit\`{a} 2-4, 08034 Barcelona, Spain
\and Landessternwarte, Zentrum für Astronomie der Universtät Heidelberg, Königstuhl 12, 69117 Heidelberg, Germany
\and Centro de Astrobiolog\'{i}a (CSIC-INTA), ESAC, Camino Bajo del Castillo s/n, 28692 Villanueva de la Ca\~{n}ada, Madrid, Spain
\and Centro Astron\'{o}mico Hispano-Alem\'{a}n (CSIC-MPG), Observatorio Astron\'{o}mico de Calar Alto, Sierra de los Filabres, 04550 G\'{e}rgal, Almer\'{i}a, Spain
\and Instituto de Astrof\'{i}sica de Canarias, V\'{i}a L\'{a}ctea s/n, 38205 La Laguna, Tenerife, Spain
\and Departamento de Astrof\'{i}sica, Universidad de La Laguna, 38206 La Laguna, Tenerife, Spain
\and Thüringer Landessternwarte Tautenburg, Sternwarte 5, 07778 Tautenburg, Germany
\and Max-Planck-Institut für Astronomie, Königstuhl 17, 69117 Heidelberg, Germany
\and Departamento de Astrof\'{i}sica y Ciencias de la Atm\'{o}sfera, Facultad de Ciencias F\'{i}sicas, Universidad Complutense de Madrid, 28040 Madrid, Spain
}

\date{Received 03 Nov 2018 / Accepted 13 Dec 2018}

\abstract
{We report the detection of a Neptune-mass exoplanet around the M4.0 dwarf GJ~4276 (G~232-070)
based on radial velocity (RV) observations obtained with the CARMENES spectrograph.
The RV variations of GJ~4276 are best explained by the presence of a planetary 
companion that has a minimum mass of $m_{\rm b}\sin i \approx 16\, M_\oplus$ on a $P_{\rm b}=13.35$ day orbit.
The analysis of the activity indicators and spectral diagnostics 
exclude stellar induced RV perturbations and 
prove the planetary interpretation of the RV signal.
We show that a circular single-planet solution can be excluded
by means of a likelihood ratio test.
Instead, we find that the RV variations can be explained either by an eccentric orbit
or interpreted as a pair of planets on circular orbits near a period ratio of 2:1.
Although the eccentric single-planet solution is slightly preferred,
our statistical analysis indicates that none of these two scenarios 
can be rejected with high confidence using the RV time series obtained so far. 
Based on the eccentric interpretation, we find that 
GJ~4276~b is the most eccentric ($e_{\rm b}=0.37$) exoplanet around an M dwarf 
with such a short orbital period known today.}


\maketitle

\section{Introduction}

M dwarfs constitute roughly 75\,\% of the stellar population 
in the solar neighborhood
\citep{Henry2006}. Compared to solar-like stars, 
they are smaller in mass, radius, and luminosity.
These properties shift the focus of ongoing and future transit and 
radial velocity (RV) surveys toward M dwarfs for many reasons.
Since the semi-amplitude of the reflex motion scales with stellar mass as 
$M_{\star}^{-2/3}$ \citep[e.g.,][]{Cumming1999}
and the transit depth with the stellar radius as $R_{\star}^{-2}$ 
\citep[e.g.,][]{Seager2003}, they are most promising targets 
for exoplanet searches and, in particular, for finding Earth-like rocky planets. 
Of special interest are planets located in the habitable zone, 
in which water can exist on the planetary surface in a liquid phase.
Due to the intrinsic faintness of M dwarfs, the distance of the habitable zone
is much smaller for those stars.
This leads
to shorter orbital periods and larger transit probabilities. 
Early M dwarfs show a high planet occurrence 
rate of $2.5\pm0.2$ planets with $1-4$ Earth radii and 
orbital periods shorter than 200 days per star \citep{Dressing2015},
implying that these objects are numerous planet hosts
in the Milky Way. 

The search for low-mass planets around a sample of about 300 M dwarfs \citep{Reiners2018_a}
is the main scientific objective of the RV survey conducted by 
the CARMENES consortium \citep{Quirrenbach2018}. 
The CARMENES instrument has already proved its ability to reach 
an RV accuracy of $\sim$1\,m\,s$^{-1}$ and has enabled the discovery 
and characterization of several planetary systems 
\citep{Trifonov2018, Reiners2018_b, Sarkis2018, Kaminski2018, Luque2018, Ribas2018}.

In this paper, we report the detection of a Neptune-mass object 
orbiting GJ~4276. 
In Sect.~\ref{host_star}, we present the stellar characteristics of GJ~4276.
The photometric data sets and the determination of the rotation period are described 
in Sect.~\ref{Photometry}.
We performed a detailed analysis of the RV measurements and the stellar activity,
and fit Keplerian models to the RV data, as described in Sect.~\ref{Spectroscopy}.
Finally, we summarize and discuss our findings in Sect.~\ref{summary}.

\section{Host star properties}
\label{host_star}
We summarize the main characteristics of our star in Table~\ref{table:stellarparameters}.
GJ~4276 (G 232-070, Karm J22252+594) is an M4.0 dwarf \citep{Reid1995, Lepine2013}
at a distance of $21.35\pm0.02$\,pc \citep{Gaia2018}.
Together with the parallax, we used the proper motion in right ascension and declination 
to calculate the secular acceleration 
($\dot \varv_{\mathrm{rad}} = 0.048\pm0.002$\,m\,s$^{-1}$\,yr$^{-1}$).
The $UVW$ Galactic space velocities 
imply that GJ~4276 belongs to the thin-disk stellar population
\citep{Cortes2016}. 

The basic photospheric parameters $T_{\mathrm{eff}}$, $\log g$, and [Fe/H]
were measured as in \citet{Passegger2018},
who fit the latest version of the PHOENIX-ACES models \citep{Husser2013} 
to CARMENES spectra. 
We computed the luminosity from the \textit{Gaia} DR2  
parallax and multiwavelength photometry from $B$ to $W4$ as described in 
\citet{Kaminski2018} and \citet{Luque2018}. 
Based on our $T_{\mathrm{eff}}$ and $L$ determinations, 
we computed the stellar radius $R$ by means of the Stefan-Boltzmann law, and 
finally derived the stellar mass $M$ using a linear mass-radius relation.
The details of the luminosity, radius, and mass determinations 
of the CARMENES targets will be presented by Cifuentes et al. (in prep.) 
and Schweitzer et al. (in prep.).

The star GJ~4276 is not a \textit{ROSAT} All-Sky Survey (RASS) source and we estimated
an upper limit for the X-ray luminosity of 
$L_X\approx 8 \times 10^{27}$\,erg\,s$^{-1}$ using the typical RASS detection limit 
of $f_X\approx2\times10^{-13}$\,erg\,cm$^{-2}$\,s$^{-1}$ \citep{Schmitt1995} 
and, from it, an upper limit of $L_\mathrm{X}/L_\mathrm{bol} < 10^{-4}$. 
According to \citet{Reiners2018_a}, GJ~4276 is not an H$\alpha$ emitter
and has an $2\,$km\,s$^{-1}$ upper limit 
on the projected rotational velocity $\varv \sin i$. 

\begin{table}
\begin{center}
\caption{Stellar parameters of GJ~4276.}
\label{table:stellarparameters}
\begin{tabular}{lll} 
\hline\hline
Parameter & GJ~4276 & Ref.\tablefootmark{a} \\ \hline
$\alpha$ & 22 25 17.32 & \textit{Gaia} DR2 \\ 
$\delta$ & +59 24 45.01 & \textit{Gaia} DR2 \\
SpT & M4.0 & Rei95, L\'{e}p13 \\
$G$ [mag] & $11.6605\pm0.0006$ & \textit{Gaia} DR2 \\
$J$ [mag] & $8.75\pm0.03$ & 2MASS\\
$\pi$ [mas] & $46.84\pm0.04$ & \textit{Gaia} DR2 \\
$\mu_{\alpha}\cos \delta$ [mas\,yr$^{-1}$] & $122.37\pm0.07$ & \textit{Gaia} DR2 \\
$\mu_{\delta}$ [mas\,yr$^{-1}$] & $-310.10\pm0.06$ & \textit{Gaia} DR2 \\
$\varv_{\rm rad}$ [km\,s$^{-1}$] & 4.034 & Rei18 \\
$U$ [km\,s$^{-1}$] & $4.4\pm0.45$ & Cor16 \\
$V$ [km\,s$^{-1}$] & $5.96\pm0.16$ & Cor16 \\
$W$ [km\,s$^{-1}$] & $ -28.46\pm1.17$ & Cor16 \\
$T_{\mathrm{eff}}$ [K]  & $3387\pm51$ & Sch18 \\
$\log\,g$ [dex]         & $4.97\pm0.07$ & Sch18 \\
$[\mathrm{Fe/H}]$ [dex] & $0.12\pm0.16$ & Sch18 \\
$M$ [$M_\odot$] & $0.406\pm0.030$ & Sch18 \\
$L$ [$L_\odot$] & $0.0197\pm0.0003$ & Sch18 \\
$R$ [$R_\odot$] & $0.407\pm0.015$ & Sch18 \\
$\varv \sin i$ [km\,s$^{-1}$] & $< 2$ & Rei18 \\
$P_{\rm rot}$ [d] & $64.3\pm1.2$ & This work \\
Age [Gyr] & $6.9\pm1.1$ & This work \\
\hline
\end{tabular}
\tablefoot{
\tablefoottext{a}{\textit{Gaia} DR2: \citet{Gaia2018};
Rei95: \citet{Reid1995};
L\'{e}p13: \citet{Lepine2013};
2MASS: \citet{Skrutskie2006};
Roe10: \citet{Roeser2010};
Rei18: \citet{Reiners2018_a};
Cor16: \citet{Cortes2016}; 
Sch18: Schweitzer et al. (in prep.)}}
\end{center}
\end{table}

\section{Photometry}
\label{Photometry}
To search for photometric modulation
caused by rotating surface inhomogeneities such as dark spots and 
bright plages,
we used archival time-series photometry from the MEarth-North project \citep{Berta2012}
and the ``All-Sky Automated Survey for Supernovae'' \citep[ASAS-SN;][]{Shappee2014}.
In addition, we obtained custom $V$ band photometry with the T150 telescope 
located at the Sierra Nevada Observatory (SNO) in Spain and with two 40\,cm telescopes 
of the Las Cumbres Observatory (LCO) located at the Haleakala Observatory on Hawai'i
and the Teide Observatory on the Canary Islands.

The MEarth-North telescope array is located at the 
Fred Lawrence Whipple Observatory, Arizona, 
and consists of eight 40\,cm robotic telescopes. 
Each is equipped with a $2048\times 2048$ CCD with a
pixel scale of $0.76''$ and a custom 715\,nm longpass filter.
While the main objective of the MEarth project is the search 
for low-mass rocky exoplanets around M dwarfs in the habitable zone
with the transit method,
ASAS-SN is dedicated to the discovery of nearby supernovae 
by monitoring the entire visible sky down to $\sim17$\,mag in the $V$ band.
It comprises five units with a total of 20 telescopes situated in 
Chile, Hawai'i, South Africa, and Texas. 
Each of the 14\,cm telephoto lenses has a 2k\,$\times$\,2k CCD 
with a field of view of 4.5\,$\times$\,4.5\,deg and 
a pixel scale of $7.8''$. 
The T150 telescope at the SNO
is a $150\,$cm Ritchie-Ch\'{e}tien telescope. It is equipped with a 2k\,$\times$\,2k VersArray CCD camera
with a field of view of 7.9\,$\times$\,7.9\,arcmin \citep{Rodriguez2010}.
The LCO telescopes are equipped with a 3k\,$\times$\,2k SBIG CCD camera 
with a pixel scale of 0.571$''$ providing a field of view of 29.2\,$\times$\,19.5\,arcmin.

The photometric measurements used in this study 
cover a time span of four years of MEarth data (October 2011 -- November 2015),
three years of ASAS-SN data (December 2014 -- December 2017),
four months of SNO data (May -- September 2018),
and three months of LCO data (June -- September 2018). 
Exposure times of ten minutes for MEarth and ASAS-SN, 
50 seconds for SNO, and 150 seconds for LCO result in median uncertainties
of $\overline{\sigma}_{\rm MEarth} = 4$\,mmag,
$\overline{\sigma}_{\rm ASAS-SN} = 15$\,mmag,
$\overline{\sigma}_{\rm SNO} = 2.9$\,mmag,
and $\overline{\sigma}_{\rm LCO} = 3.2$\,mmag.

\subsection{Rotation period analysis}
To identify potentially spot induced periodic variability,
we applied the generalized Lomb-Scargle (GLS) periodogram 
\citep{Zechmeister2009_b}
to the MEarth, ASAS-SN, and SNO data sets of GJ~4276.
The periodograms show evidence for periodicity at 
$P_{\mathrm{MEarth}} = 63.9^{+1.7}_{-1.6}\,$d, $P_{\mathrm{ASAS-SN}} = 64.7^{+1.6}_{-1.5}\,$d,
and $P_{\rm SNO} = 32.3^{+4.3}_{-3.4}\,$d.
To estimate the uncertainties of our period determination,
we fit a Gaussian profile to the peak with the largest power 
and computed its full-width-half-maximum (FWHM).

\begin{figure}
\begin{center}
\includegraphics[width=0.5\textwidth]{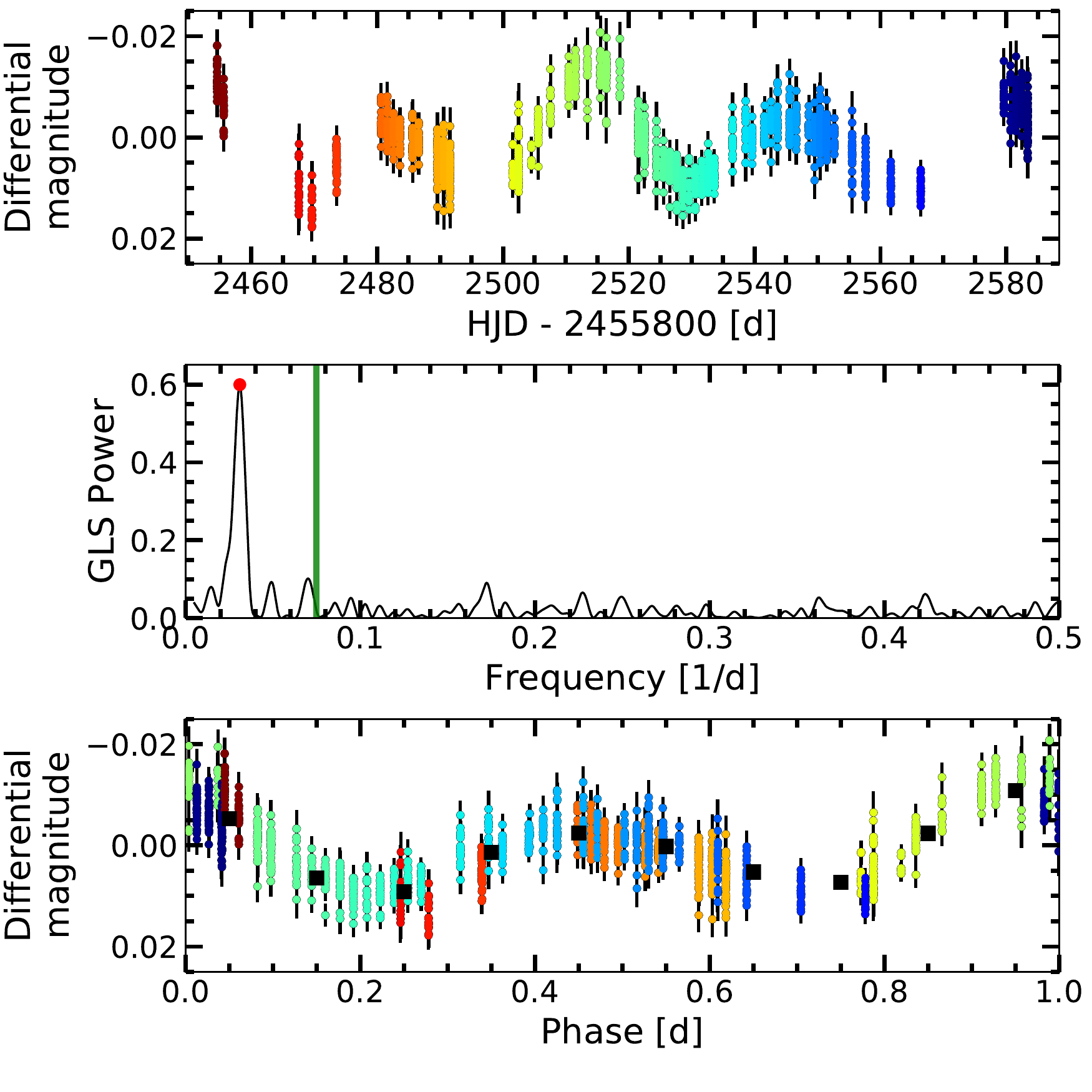}
\caption{Rotation period analysis using SNO photometric data.
\textit{Top:} $V$ band light curve. The color of the datapoints indicates 
the observation epoch. 
\textit{Middle:} GLS periodogram. The vertical green line
represents the orbital period of the planet at 13.35 days and the red dot
the peak with the highest power at 32.3 days. 
\textit{Bottom:} Phased light curve using twice the period derived from the GLS.
The black squares indicate the mean magnitude in ten equidistant bins in phase.}
\label{figure:SNO}
\end{center}
\end{figure}

We present the light curve, the periodogram,
and the phase folded light curve derived from the SNO data in Fig.~\ref{figure:SNO}.
The light curves and periodograms of the MEarth and ASAS-SN data sets
are shown in Fig.~\ref{figure:photometry_appendix}. 
Visual inspection of the SNO light curve (top panel of Fig.~\ref{figure:SNO})
shows a clear variability pattern, 
which remained rather stable during the observation run. The pattern is well resolved
and consists of two bumps with alternating amplitude, which we
interpret as the photometric manifestation of two starspots located on opposing
hemispheres. Therefore, we conclude that
the GLS peak at $32.3$ days is the semi-period of the stellar rotation period
of $\approx 64.6$\,d, which also resolves the apparent conflict with the MEarth and ASAS-SN
data. The phase-folded light curves 
of the latter show a less pronounced signal, which may be related to the longer span
covered. We find consistent results with the LCO data.

The rotation period obtained here is consistent with the findings of
\citet{DiezAlonso2018}, who 
reported a value of $64.6\,\pm2.1\,$d with a FAP level of $<10^{-4}\,\%$
for GJ~4276 based on their analysis of the ASAS-SN light curve alone.
Also, a rotation period of roughly 64 days 
is consistent with the low activity level observed in GJ~4276 
and the absence of H$\alpha$ emission.
Based on gyrochronological models by \citet{Barnes2007}, 
we calculated an age of $6.9\pm1.1\,$Gyr using the intrinsic $B-V$ color 
and the derived rotation period as input parameters.  

\begin{figure}
\begin{center}
\includegraphics[width=0.5\textwidth]{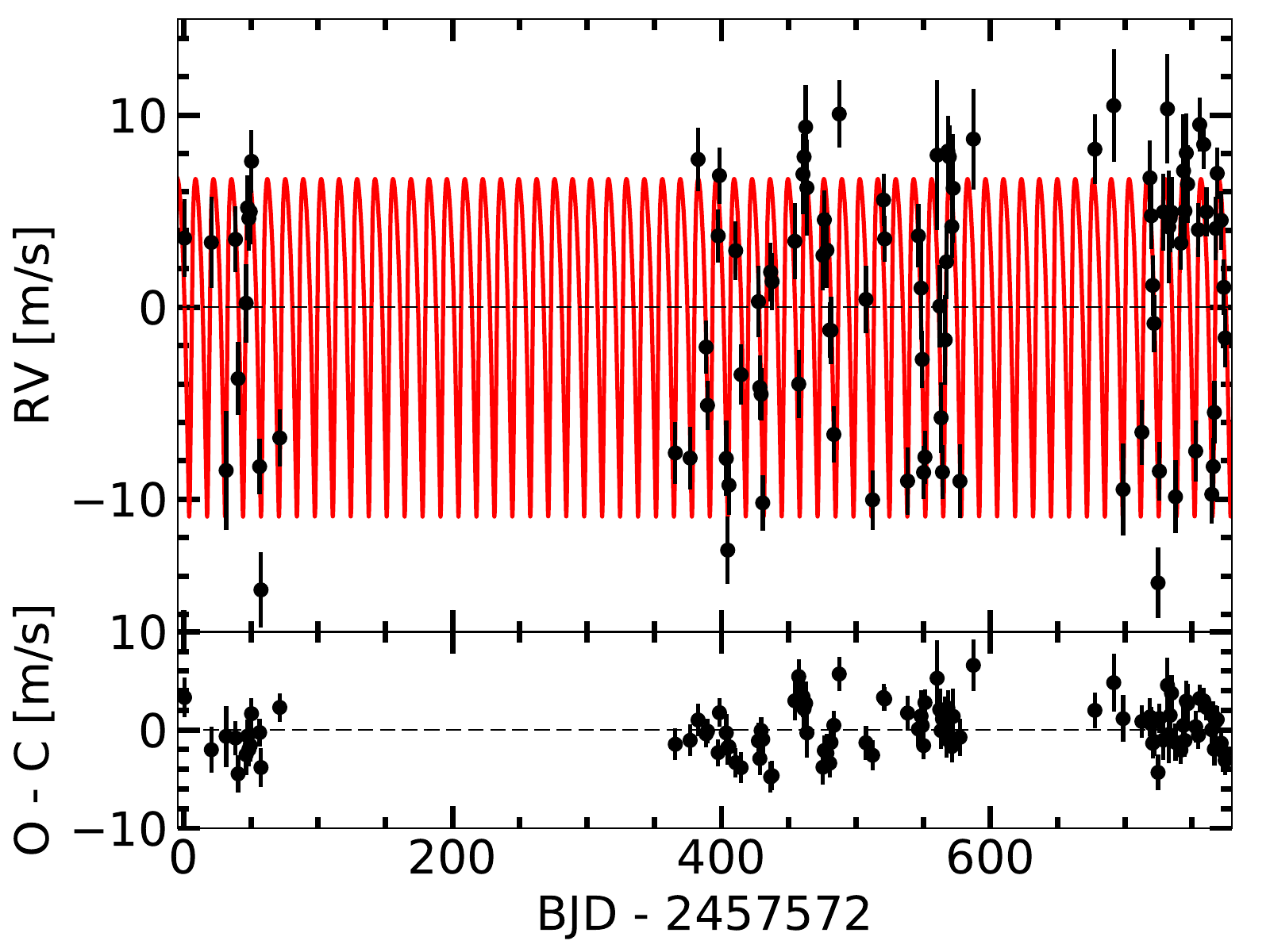}
\caption{\textit{Top:} Radial velocity measurements of GJ~4276 obtained with CARMENES 
as a function of barycentric Julian Date. The best-fit eccentric single-planet Keplerian model
is overplotted in red (see Sect.~\ref{subsection:orbitalSolution}).
\textit{Bottom:} $\mathrm{O} - \mathrm{C}$ residuals.}
\label{figure:rvFit}
\end{center}
\end{figure}

\begin{figure}
\begin{center}
\includegraphics[width=0.49\textwidth]{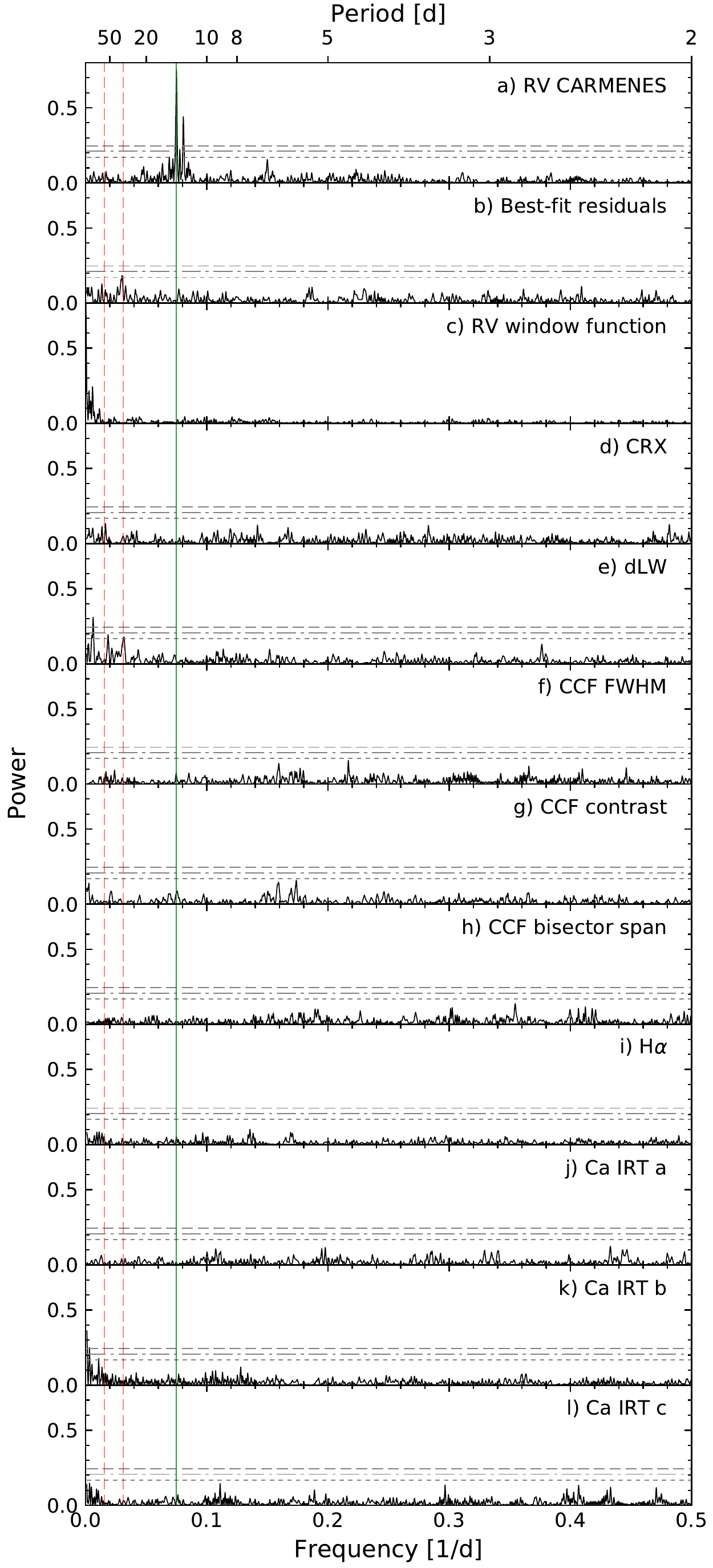}
\caption{GLS periodograms of GJ~4276. 
The periodogram of the CARMENES RVs is shown in (a). 
The horizontal lines (dotted, dash-dotted, dashed) 
indicate FAP levels of 10\,\%, 1\,\%, and 0.1\,\%. 
The vertical green line 
marks the orbital period with the highest power at 13.347\,d.
The left red dashed line at $P_{\mathrm{rot}} = 64.3$\,d
($f_{\mathrm{rot}} = 0.0156\,$d$^{-1}$) shows 
the weighted mean of the photometrically derived stellar rotation period
and the right red dashed line 
its first harmonic ($2f_{\mathrm{rot}} = 0.0311\,$d$^{-1}$).
The periodogram of the residuals after removing the best-fit 
single-planet Keplerian with eccentricity 
(see Sect.~\ref{subsubsection:single_eccentric_keplerian})
signal is shown in (b) and the window function of the RV data in (c). 
Panels (d-h) show periodograms of the chromatic RV index (CRX), 
differential line width (dLW), as well as FWHM, contrast, and bisector span
from the CCF analysis.  
Periodograms of the chromospheric line indices of H$\alpha$ and \ion{Ca}{II} IRT 
are presented in (i-l).}
\label{figure:periodogram}
\end{center}
\end{figure}

\section{Spectroscopy}
\label{Spectroscopy}
We gathered exactly 100 CARMENES RV measurements of GJ~4276
over a time span of 774 days.
The observations were carried out 
as part of the CARMENES GTO survey \citep{Reiners2018_a}
between July 2016 and August 2018 with the CARMENES
echelle spectrograph \citep{Quirrenbach2018}, 
mounted on the 3.5\,m telescope of the 
Calar Alto Observatory in Spain. 
CARMENES consists of a pair of high resolution 
spectrographs, which cover the optical wavelength range 
from $5200\,\AA$ to $9600\,\AA$ with a resolution power of $R=94\,600$,
and the near-infrared range from $9600\,\AA$ to $17\,100\,\AA$ 
with $R=80\,400$.
Both channels are enclosed in temperature- and pressure-stabilized 
vacuum vessels to reduce instrumental drifts and to provide 
a RV precision on a m\,s$^{-1}$ level. 

The CARMENES survey observation strategy aims at reaching a 
signal-to-noise ratio of $150$ in the $J$ band.
The typical exposure time of our spectra of GJ~4276
is 1800\,s. 
The raw frames were extracted using the CARACAL reduction pipeline
\citep{Caballero2016_2}, which is based on flat-relative optimal 
extraction \citep{Zechmeister2014}. The wavelength calibration is based 
on three hollow cathode lamps (U-Ne, U-Ar, and Th-Ne) combined with a 
Fabry-P\'{e}rot etalon \citep{Bauer2015, Schaefer2018}. The reference frames were taken at the 
beginning of each observing night. In addition, Fabry-P\'{e}rot etalon spectra 
were taken simultaneously with the target to track and correct the nightly 
instrument drift.

To precisely measure the Doppler shifts on a m\,s$^{-1}$ level, 
we used the SERVAL\footnote{SpEctrum Radial Velocity AnaLyser, \\
\url{https://github.com/mzechmeister/serval}} code \citep{Zechmeister2018},
which constructs a high signal-to-noise template spectrum by coadding 
all spectra of GJ~4276 after correcting for barycentric motion
\citep{Wright2014} and secular acceleration \citep{Zechmeister2009_a}. 
To consider systematic instrumental effects, 
we further corrected the RVs for nightly zero-point variations using 
RV measurements of stars with low RV variability observed in the same night;
we refer to \citet{Trifonov2018} for a detailed description.

In this study, we employed RVs only from the VIS channel,
which have an internal median uncertainty of 1.7\,m\,s$^{-1}$.
We present the RV measurements used in this paper in Fig.~\ref{figure:rvFit}
and list them along with their formal 
uncertainties in Table~\ref{appendix:rvs}.

\subsection{Periodogram analysis}
To study the RV variability of GJ~4276,
we applied the GLS periodogram 
to the measurements obtained with CARMENES. The resulting periodogram is shown in
Fig.~\ref{figure:periodogram}. 
Following Eq.~24 from \citet{Zechmeister2009_b}, we computed the false alarm probabilities (FAPs)
to evaluate the significance of the peaks in the power spectra.

The largest power excess with a FAP well below $0.1\,\%$ appears at a frequency
of $f=0.07493$\,d$^{-1}$ (13.347 days, Fig.~\ref{figure:periodogram}a). 
To check the persistence of this signal, we
divided the entire data set into 
three RV subsamples and separately analyzed their periodograms. In all cases
we find similar peaks, corresponding to frequencies of
$f_1=0.07480\,$d$^{-1}$ (13.370 days), 
$f_2=0.07438\,$d$^{-1}$ (13.444 days), and $f_3=0.07581\,$d$^{-1}$ (13.192 days),
indicating that the signal is, indeed, persistent.
Furthermore, a power peak of the first harmonic of the dominant signal at 
$f=0.14986$\,d$^{-1}$ (6.673 days) is visible in the periodogram.

We identify further strong signals with $\mathrm{FAPs} < 0.1\,\%$ at frequencies of
$0.92784$\,d$^{-1}$ and $1.07766\,$d$^{-1}$ 
with powers of $0.51$ and $0.58$, respectively
(outside the frequency range shown in Fig.~\ref{figure:periodogram} for the sake of clarity). 
Both peaks are plausible one-day aliases of the primary period 
($\sim 1.000\pm 0.075$\,d$^{-1}$)
which disappear after we subtract 
the best-fit eccentric single-planet Keplerian model 
(see Sect.~\ref{subsubsection:single_eccentric_keplerian})
from the RV measurements. 

To ensure that the RV variation is not caused by stellar activity,
we made use of several spectral diagnostics
provided by SERVAL, viz., chromospheric indices, the differential line width,
the chromatic index, and the cross-correlation function.
The chromatic index (CRX), as introduced by \citet{Zechmeister2018},
describes the color-dependence of the RV signal, which must vanish
for a planetary signal but not for a spot-induced signal.
Rotating spots induce periodic line profile variations, which were
scrutinized using the differential line width (dLW) indicator. 
We also analyzed the cross-correlation function (CCF)
of each spectrum. Specifically, we checked for periodic modulation of the
FWHM, contrast, and bisector span
as described in \citet{Reiners2018_b}. Any such detection would, again,
be a red flag indicating activity-induced modulation. Finally, 
the H$\alpha$ and \ion{Ca}{II} IRT line indices were analyzed, which
directly trace chromospheric activity.

We present GLS periodograms of all these spectral diagnostic time series 
in Fig.~\ref{figure:periodogram}. 
Beside the periodogram of the dLW and \ion{Ca}{II} IRT b line indices at $8542\,\AA$,
none of the investigated indicators exhibit significant peaks above the 10\,\% FAP level.
The periodogram of the dLW shows a marginally significant power peak 
at the first harmonic of the stellar rotation period around 32 days, which is 
most likely caused by rotational modulation of active regions.
In addition, the dLW shows two peaks at 54 and 161 days with formal FAP levels above 10\,\%
and 0.1\,\%, respectively.
Also, some long-term periodic pattern can be seen in the periodogram
of the \ion{Ca}{II} IRT b line index but no peaks were found 
in the periodograms of the \ion{Ca}{II} IRT a and c line indices 
at $8498\,\AA$ and $8662\,\AA$ at similar frequencies. 
Importantly, however, the RV signal at 13.347 days correlates neither with 
the spurious signals produced by the dLW and the \ion{Ca}{II} IRT b line indices
nor with any signal produced by other spectral activity indicators. 
Thus, we are confident that this persistent signal is not related to activity, 
but is most probably of planetary origin.

\subsection{Orbital solutions}
\label{subsection:orbitalSolution}

\begin{figure}
\begin{center}
\includegraphics[width=0.5\textwidth]{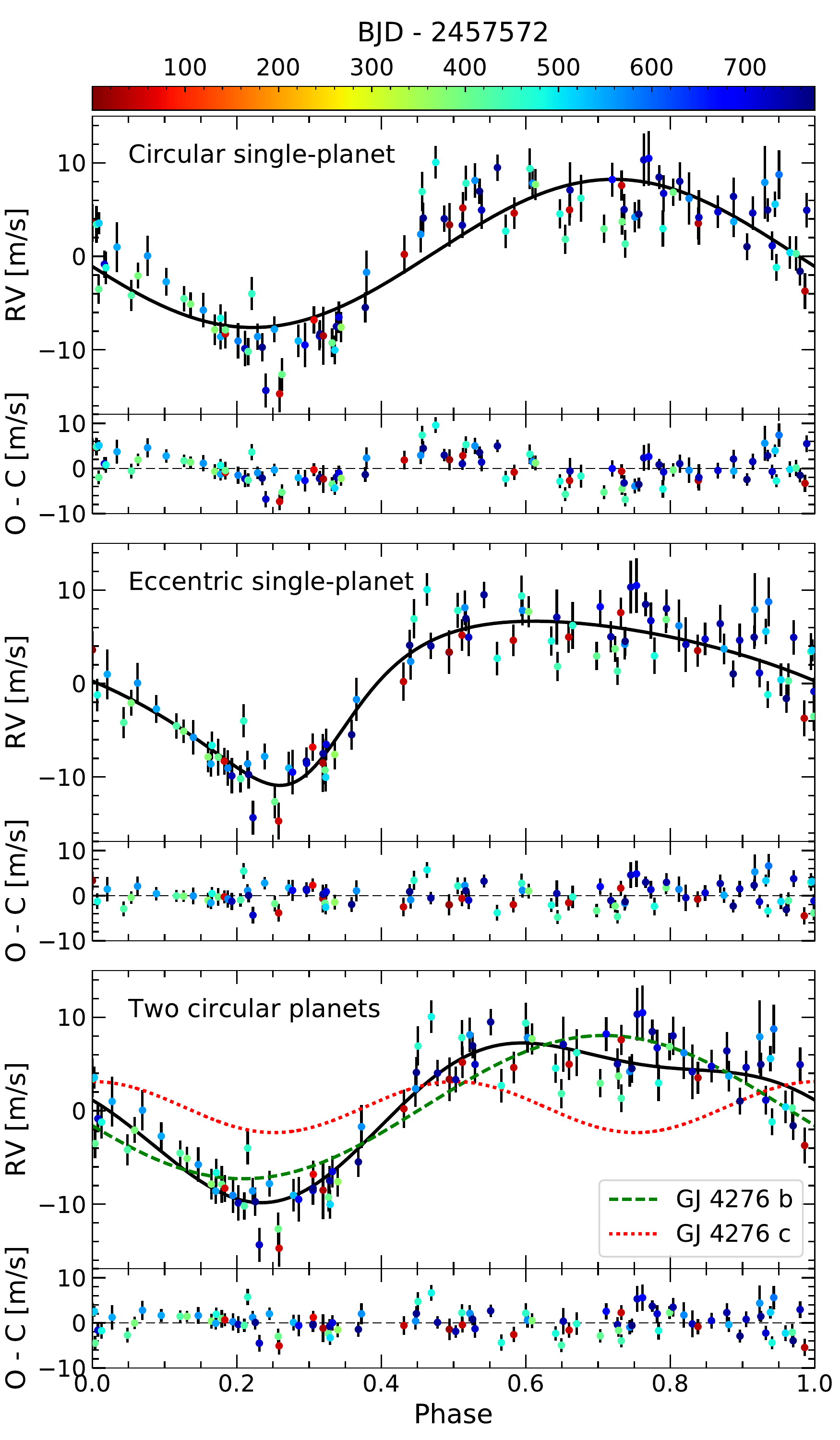}
\caption{Phase-folded radial velocity measurements of GJ~4276,
together with the best-fit Keplerian model (black line) overplotted. 
In the bottom of each panel we show the $\mathrm{O} - \mathrm{C}$ residuals.
\textit{Top:} Circular single-planet Keplerian model.
\textit{Middle:} Eccentric single-planet Keplerian model.
\textit{Bottom:} Two-planet Keplerian model on circular orbits with a period ratio of 2:1. 
In addition, we show the best-fit Keplerian model of GJ~4276~b (green dashed line) 
and GJ~4276~c (red dotted line).}
\label{figure:rvFitFolded}
\end{center}
\end{figure}

\begin{table*}
\begin{center}
\caption{Best-fit orbital parameters for the GJ~4276 system.}
\label{table:orbitalparameters}
\begin{tabular}{lcccccrcccccrcccccrrccc} 
\hline\hline
\Tstrut Orbital parameters &&&&&& GJ~4276~b\tablefootmark{(1)} &&&&&& GJ~4276~b\tablefootmark{(2)} &&&&&& GJ~4276~b\tablefootmark{(3)} & GJ~4276~c\tablefootmark{(3)}\\
\hline 
\Tstrut\Bstrut $K$ [m\,s$^{-1}$] &&&&&& $7.93^{+0.32}_{-0.32}$ &&&&&& $8.79^{+0.27}_{-0.27}$ &&&&&& $7.67^{+0.25}_{-0.25}$ & $2.73^{+0.23}_{-0.24}$ \\
\Tstrut\Bstrut $P$ [d] &&&&&& $13.348^{+0.005}_{-0.005}$ &&&&&& $13.352^{+0.003}_{-0.003}$ &&&&&& $13.350^{+0.004}_{-0.004}$ & $6.675^{+0.002}_{-0.002}$ \\
\Tstrut\Bstrut $e$ &&&&&& 0 (fixed) &&&&&& $0.37^{+0.03}_{-0.03}$ &&&&&& $0$ (fixed) & $0$ (fixed)\\ 
\Tstrut\Bstrut $\omega$ [deg] &&&&&& $90$ (fixed) &&&&&& $216.83^{+4.74}_{-4.56}$ &&&&&& $90$ (fixed) & $90$ (fixed) \\ 
\Tstrut\Bstrut $\tau$ [BJD $ - 2457572$] &&&&&& $0.28^{+0.23}_{-0.22}$ &&&&&& $4.72^{+0.19}_{-0.17}$ &&&&&& $0.10^{+0.17}_{-0.17}$ & $2.35^{+0.18}_{-0.18}$ \\ 
\Tstrut\Bstrut $\gamma$ [m\,s$^{-1}$] &&&&&& $0.31^{+0.24}_{-0.23}$ &&&&&& $0.52^{+0.18}_{-0.18}$ &&&&&& \multicolumn{2}{c}{$0.39^{+0.18}_{-0.17}$} \\
\Tstrut\Bstrut $\sigma_{\mathrm{jitter}}$ [m\,s$^{-1}$] &&&&&& $2.83^{+0.22}_{-0.20}$ &&&&&& $1.74^{+0.18}_{-0.17}$ &&&&&& \multicolumn{2}{c}{$1.89^{+0.18}_{-0.17}$} \\
\Tstrut\Bstrut $a$ [au] &&&&&& $0.082^{+0.002}_{-0.002}$ &&&&&& $0.082^{+0.002}_{-0.002}$ &&&&&& $0.082^{+0.002}_{-0.002}$ & $0.051^{+0.001}_{-0.001}$ \\ 
\Tstrut\Bstrut $m_{\mathrm{p}}\sin i$ [M$_\oplus$] &&&&&& $16.11^{+1.03}_{-1.01}$ &&&&&& $16.57^{+0.94}_{-0.95}$ &&&&&& $15.58^{+0.93}_{-0.90}$ & $4.40^{+0.44}_{-0.44}$ \\ 
\Tstrut\Bstrut $\sigma_{\mathrm{O-C}}$ [m\,s$^{-1}$] &&&&&& 3.29 &&&&&& 2.46 &&&&&& \multicolumn{2}{c}{2.57} \\
\Tstrut\Bstrut $-2\ln \mathcal{L} $ &&&&&& $229.71$ &&&&&& $170.48$ &&&&&& \multicolumn{2}{c}{179.77} \\
\hline 
\end{tabular}
\tablefoot{
\tablefoottext{1}{Circular single-planet Keplerian model,}
\tablefoottext{2}{eccentric single-planet Keplerian model,}
\tablefoottext{3}{two-planet Keplerian model on circular orbits with period ratio of 2:1.}
}
\end{center}
\end{table*}

Having established the planetary origin of the RV signal, we now
determine the orbital elements of the planet. 
To that end, we implemented a Keplerian RV curve model and carry out
parameter optimization using a 
Nelder-Mead simplex algorithm \citep{Nelder1965}. 
Following the approach of \citet{Baluev2009}, our model incorporates an 
RV jitter variance term to account for additional stochastic scatter;
the jitter parameter is fit simultaneously during the parameter optimization.

In the following, we juxtapose three Keplerian models and their performance in
describing the observations, in particular,
a single planet with a circular orbit, a single planet with an eccentric orbit,
and two planets with circular orbits 
with a period ratio of 2:1.
In addition to the periodic planetary signal, we allow for an RV offset
to account for entire system's velocity.
In a first run, we further fit a linear time evolution parameter to derive 
potential systematic acceleration and gauged with a likelihood ratio test 
whether the improvement is sufficient to include a slope. 
Although this additional fitting parameter resulted in a higher likelihood, 
we find that the improvement is non-significant. 

The entire set of the derived best-fit Keplerian orbital elements 
is displayed in Table~\ref{table:orbitalparameters}.
The $1\sigma$ uncertainties of the 
orbital parameters are estimated from the posterior distributions 
using the Markov Chain Monte Carlo sampler \url{emcee} 
\citep{Foreman-Mackey2013} along with our Keplerian models 
(Figs.~\ref{figure:corner1}-\ref{figure:corner3}).
For the fit parameters we assumed uniform priors, except for the stellar mass,
for which we imposed a Gaussian prior with mean and variance 
equal to $0.406\pm0.030\,M_\odot$ 
based on the mass determination of GJ 4276 (Sect.~\ref{host_star}).

\subsubsection{Single planet on circular orbit}
In the first model, we fit the RV measurements with a single
planet on a circular orbit. 
We left the semi-amplitude $K_{\mathrm{b}}$,
the orbital period $P_{\mathrm{b}}$,
the RV jitter $\sigma_{\mathrm{jitter}}$,
as well as the RV offset $\gamma$ 
as free parameters. The eccentricity remained fixed to $e_{\mathrm{b}}=0$. 
Further, we also fixed the argument of the periapsis to $\omega_{\mathrm{b}}=90\,$deg
and fit the time of the periastron passage $\tau_{\rm b}$.

This model converges on a period of $P_{\mathrm{b}}=13.348$ days,
matching the frequency of the power peak found in the periodogram.  
Following a Keplerian interpretation of the RV variations, 
GJ~4276~b is a Neptune-like planet with a minimum mass of 
$m_{\mathrm{b}} \sin i = 16.11$\,M$_{\oplus}$. 
Orbiting at a distance of 0.082\,au from its host star, 
it is placed closer than the inner edge of the conservative and 
optimistic habitable zones, 
which range from 0.146 to 0.284\,au 
and 0.115 to 0.299\,au, respectively \citep{Kopparapu2013, Kopparapu2014}. 
The solution further yields a semi-amplitude of $K_{\mathrm{b}}=7.93\,$m\,s$^{-1}$
and a jitter term of $\sigma_{\mathrm{jitter}} = 2.83\,$m\,s$^{-1}$. 
With respect to the model, the data yield a root mean square (rms) value of 
$\sigma_{\mathrm{O-C}} = 3.29$m\,s$^{-1}$.

\subsubsection{Single planet with eccentric orbit}
\label{subsubsection:single_eccentric_keplerian}
In addition to the parameters from the circular solution, 
we here let the eccentricity $e$ and 
the argument of the periapsis $\omega_{\mathrm{b}}$ vary freely.
While the best-fit minimum mass, the orbital period, semi-amplitude, and semi-major axis
are comparable to that of the single-planet circular solution 
($m_{\mathrm{b}} \sin i = 16.57$\,M$_{\oplus}$, $P_{\mathrm{b}}=13.352$\,d, 
$K_{\mathrm{b}}=8.79\,$m\,s$^{-1}$, $a_{\mathrm{b}}= 0.082\,$au),
the jitter term of
$\sigma_{\mathrm{jitter}} = 1.74\,$m\,s$^{-1}$ found here is $1.09\,$m\,s$^{-1}$ smaller than that
previously obtained. 
The introduction of the eccentricity $e_{\mathrm{b}}=0.37$ significantly improves the fit 
and results in an rms of $\sigma_{\mathrm{O-C}} = 2.46$m\,s$^{-1}$.
We show the phased RV data and the best-fit Keplerian one-planet solution  
in Fig.~\ref{figure:rvFitFolded}.

\subsubsection{Two planets on circular orbits with period ratio of 2:1} 

The single-planet model with an eccentric orbit results in a remarkably
eccentric orbit with $e=0.37$. 
Since the Doppler signal of a two-planet system on circular orbits near a 2:1 
mean motion resonance can be misinterpreted as an eccentric single-planet
\citep{Anglada2010, Wittenmyer2013, Kuerster2015, Boisvert2018},
we further tried to fit a two-planet model with circular orbits
and fixed period ratio of 2:1, i.e., $P_{\mathrm{b}}=2P_{\mathrm{c}}$.
In the modeling, we leave $K_{\mathrm{b}}$, $K_{\mathrm{c}}$, 
$P_{\mathrm{b}}$, $\tau_{\mathrm{b}}$, $\tau_{\mathrm{c}}$,
$\gamma$, and $\sigma_{\mathrm{jitter}}$ free to vary
(whereas $\omega_{\rm b}= \omega_{\rm c}= 90\,$deg).

Based on this double Keplerian model, we obtained orbital parameters for GJ~4276~b:
$K_{\mathrm{b}} = 7.67$\,m\,s$^{-1}$, $P_{\mathrm{b}}=13.350$ days,
and for GJ~4276~c:
$K_{\mathrm{c}} = 2.73$\,m\,s$^{-1}$, $P_{\mathrm{c}}=6.675$ days, 
which translates into
minimum planetary masses of $m_{\mathrm{b}}\sin i =15.58\,M_{\oplus}$ and 
$m_{\mathrm{c}}\sin i =4.40\,M_{\oplus}$ and semi-major axes of
$a_{\mathrm{b}}=0.082$\,au and $a_{\mathrm{c}}=0.051$\,au.

\subsection{Likelihood analysis}
\label{sec:stat}

To compare the fit qualities between the eccentric single-planet model 
and the two-planet model compared to the circular single-planet model, 
we carried out likelihood ratio tests \citep[e.g.,][]{Wilks1938, Protassov2002}.
In our circular single-planet model we have five free parameters, while there are
seven in both the single-planet model with elliptical orbit and our two-planet model.
The test statistic is $-2\Delta \ln \mathcal{L}$.
According to Wilk's theorem \citep{Wilks1938}, 
the probability distribution of the test statistic can be 
approximated by a $\chi^2$ distribution with $df$ degrees of freedom
for large data samples. However,
as discussed by \citet{Protassov2002}, \citet{Baluev2009}, and \citet{Czesla2010},
the formal criteria for this approximation are not fulfilled in the
current case. While the models are nested as required, the circular single-planet
model is only obtained from our elliptical or two-planet models by choosing
parameters at the edge of the parameter space such as zero eccentricity.

\begin{figure}
\begin{center}
\includegraphics[width=0.5\textwidth]{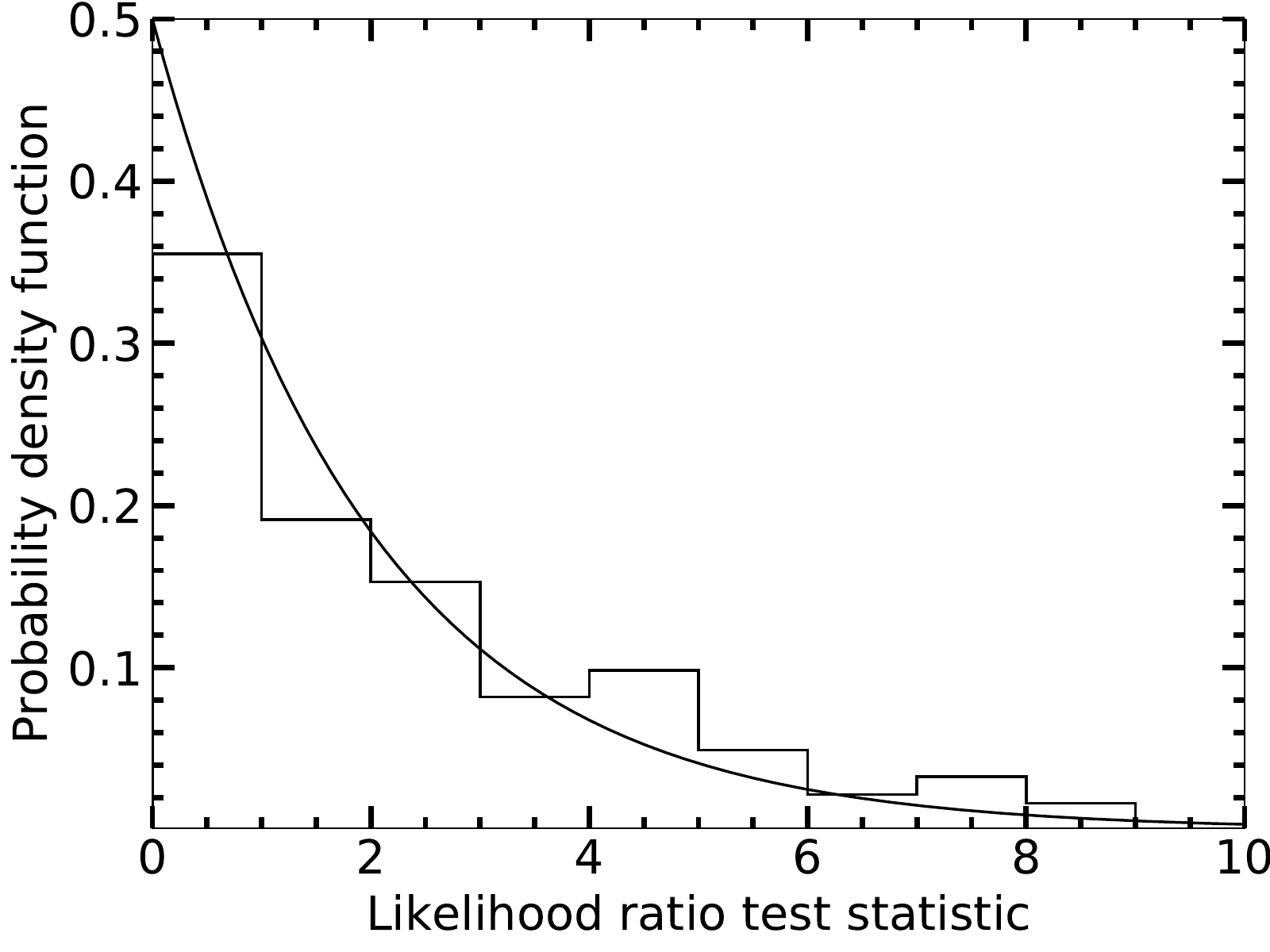}
\caption{Empirical distribution of the $-2\Delta \ln \mathcal{L}$ statistic
(histogram) along with the probability density of the $\chi^2$ distribution with two
degrees of freedom (black curve).
}
\label{figure:LRT}
\end{center}
\end{figure}

Therefore, we verified that the probability distribution of the test statistic
can, indeed, be approximated by a $\chi^2$ distribution with two degrees of freedom:
\begin{equation}
df=df_{\rm alternative} - df_{\rm null} = 8-6=2.
\end{equation}
Based on the best-fit 
circular single-planet solution, 
we generated 1000 synthetic data sets with random normally distributed errors 
that include the measurement error and the maximum-likelihood estimate of the
stellar jitter so that
$\sigma_i^2 = \sigma_{\mathrm{meas,}i}^2 + \sigma_{\mathrm{jitter}}^2$.
We fit these mock data sets using the circular single-planet model, 
as well as the eccentric single-planet and two-planet models.
Based on the maxima of the respective likelihood functions, we calculated
the test statistic $-2 \Delta \ln \mathcal{L}$. 
As an example, we show the simulated distribution of the likelihood ratio test
statistic, as well as the 
$\chi^2$ distribution, for the comparison of the circular single-planet model 
and the eccentric single-planet model in Fig.~\ref{figure:LRT}. Based on our
simulations, we conclude that the $\chi^2$ distribution yields an acceptable
approximation to the distribution of the test statistic in our case.

To assess the fit quality of the eccentric single-planet model 
compared to the circular single-planet model, we computed the ratio of the best-fit
likelihoods for the circular model $\ln \mathcal{L}_{\mathrm{1cp}}$ and eccentric model
$\ln \mathcal{L}_{\mathrm{1ep}}$ and found a value of 
$\ln \mathcal{L}_{\mathrm{1ep}} - \ln \mathcal{L}_{\mathrm{1cp}} = 29.62$.
The probability to obtain such an improvement by chance if the true orbit were circular
is only $1.4\times10^{-13}$.
The comparison between the circular single-planet scenario and the two-planet model
results in a likelihood ratio of 
$\ln \mathcal{L}_{\mathrm{2cp}} - \ln \mathcal{L}_{\mathrm{1cp}} = 24.97$, where 
$\ln \mathcal{L}_{\mathrm{2cp}}$ is the best-fit likelihood of the circular two-planet model.
Again, we find a probability of only $1.4\times10^{-11}$ that such an improvement in fit
quality can be achieved by chance. We therefore conclude
that the circular single-planet solution
can be rejected with high confidence.

\begin{figure}
\begin{center}
\includegraphics[width=0.5\textwidth]{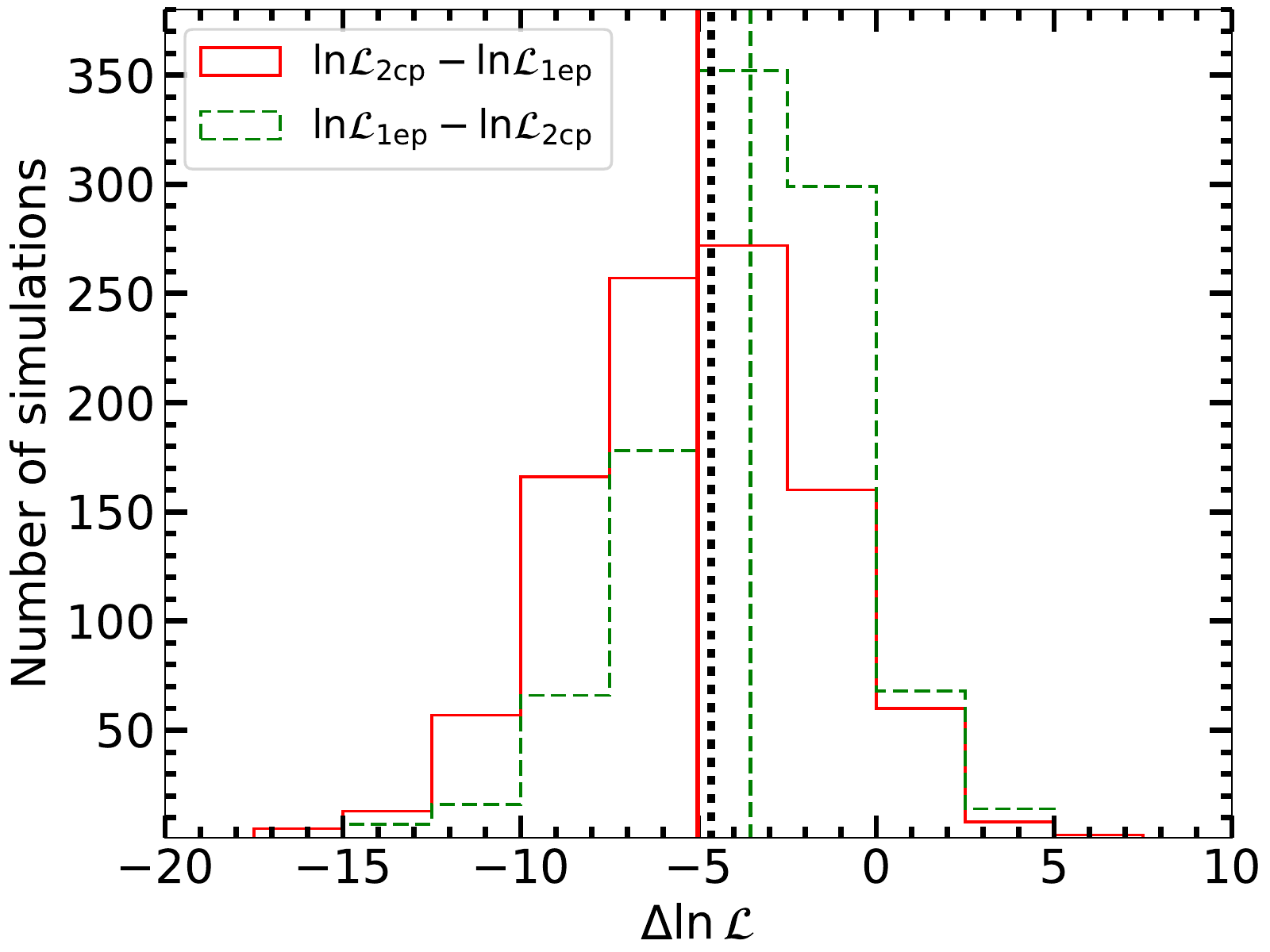}
\caption{Histograms of the likelihood ratio using simulated data sets 
based on the best-fit eccentric single-planet model (solid red) and 
the circular two-planet model (dashed green). 
The vertical solid red line and the dashed green line represent the median values of the histograms. 
The black dotted vertical line represents the measured likelihood ratio 
$\ln \mathcal{L}_{\mathrm{2cp}} - \ln \mathcal{L}_{\mathrm{1ep}} = -4.65$.}
\label{figure:guillem}
\end{center}
\end{figure}

To study whether
the eccentric single-planet model or the circular two-planet model
is statistically preferred, we carried out another simulation.
In particular, we generated 1000 artificial data sets by adding normally
distributed random noise to the
maximum-likelihood eccentric single-planet model on the one hand
and the two-planet model on the other hand.
To determine what differences in likelihood can be expected,
we fit all of these artificial RV curves using both the
eccentric single-planet and the two-planet model
and calculated the likelihood ratios
$\ln \mathcal{L}_{\mathrm{2cp}} - \ln \mathcal{L}_{\mathrm{1ep}}$
and $\ln \mathcal{L}_{\mathrm{1ep}} - \ln \mathcal{L}_{\mathrm{2cp}}$, respectively.
In Fig.~\ref{figure:guillem} we show the resulting histograms of the 
likelihood ratios. 
We find a median value of $-5.02$ assuming that the eccentric model is true
and $-3.54$ for the two-planet case. 
In addition, we indicate the measured likelihood ratio of 
$\ln \mathcal{L}_{\mathrm{2cp}} - \ln \mathcal{L}_{\mathrm{1ep}} = -4.65$. 
Based on the higher likelihood achieved in the fit, we find a slight
preference for the eccentric single-planet solution. However, our findings
show that the measured difference in likelihood does not allow to reject
one or the other solution with reasonable confidence.

One possible strategy to discriminate between the two degenerated models 
is to increase the number of RV measurements, as suggested by 
\citet{Anglada2010}, \citet{Kuerster2015}, and \citet{Boisvert2018}. 
Ideally, the observations should be carried out at phases of maximal differences
between the models. 
In the case of GJ~4276~b, we find a maximal difference of 2.10\,m\,s$^{-1}$, which
lies above the internal median error of 1.7\,m\,s$^{-1}$.
However, even for a quiet star like GJ~4276, 
we found an activity-induced RV jitter level in the range of 
1.5--3\,m\,s$^{-1}$ limiting the achievable RV accuracy. 
To provide a rough estimate on the amount of additional RV observations
that are necessary to distinguish between the two solutions,
we generated synthetic RV measurements based on the best-fit eccentric model
and fit them with the eccentric and the two-planet Keplerian model. 
Our results imply that $\sim 100$ additional measurements randomly distributed in phase 
would be sufficient to push the likelihood ratio to 
$\ln \mathcal{L}_{\mathrm{2cp}} - \ln \mathcal{L}_{\mathrm{1ep}} \approx -15$.

\subsection{Orbital evolution of the eccentric single-planet solution}
We employed an estimate of the tidal circularization timescale for the
eccentric one-planet solution in order to assess its plausibility
compared to the two-planet solution. Following
\citet{Jackson2008}, we solved the coupled differential
equation for the evolution of the semi major axis and eccentricity due
to tidal interaction. The two parameters determining this evolution
are the modified tidal dissipation values $Q$. Here we adopted
$Q_\star =10^5$ for the star. For the planet we used $Q_{\rm p} = 100$ for a possible rocky planet 
and $Q_{\rm p} = 10^5$ for a Neptune-like planet. Due to the significantly
higher dissipation, a rocky planet's orbit would completely circularize
within $10^8$\,yr, while a Neptune-like planet would maintain a high eccentricity
over more than 10 Gyr. The unknown planetary interior therefore does not
allow to provide an additional constraint to distinguish between the two 
configurations.

\subsection{Search for additional planetary companions}

To check whether the RV data yield evidence for additional planets,
we removed the best-fit single-planet eccentric model and the 
circular two-planet model from the RV data and 
investigated the GLS periodograms of the RV residuals. 
Both periodograms show power excess at 32 days on a 10\,\% FAP level, 
reflecting half of the stellar rotation period
(see Fig.~\ref{figure:periodogram}b). 
In addition to that, the periodograms of the RV residuals did not reveal 
any further significant power peaks attributable to 
planetary companions.

\section{Summary and discussion}
\label{summary}

In this study, we analyzed 100 RV measurements of 
the M4.0V star GJ~4276,
taken with the visible channel of the high-resolution CARMENES
echelle spectrograph.
The rotation period of 64 days determined from 
long-term photometry (MEarth, ASAS-SN) and the photometric campaign 
carried out during the present work (SNO, LCO),
together with the lack of H$\alpha$ emission, 
implies that GJ~4276 is a weakly active and slowly rotating star. 
The examination of the spectral diagnostics 
and the activity indicators revealed no link between 
stellar activity and the supposed planetary signal 
supporting the fact that the RV variation at this period 
arises from Keplerian motion of a planetary companion. 

\begin{figure}
\begin{center}
\includegraphics[width=0.5\textwidth]{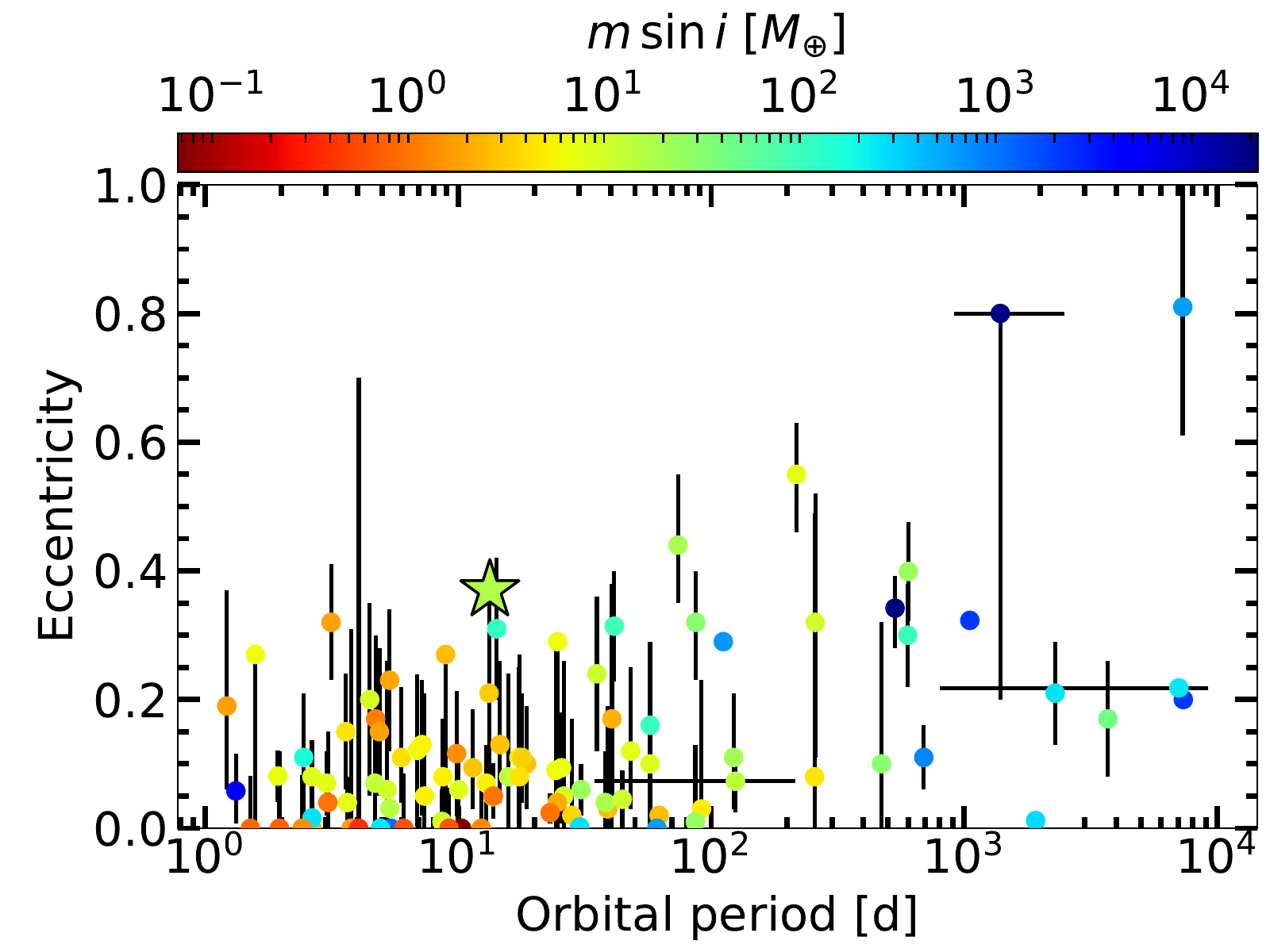}
\caption{Eccentricity plotted against orbital period of 
known exoplanets around M dwarfs (dots). 
The colors indicate the minimum mass and the star marks the position of GJ~4276~b
with the eccentric single-planet solution.}
\label{figure:eccentricities}
\end{center}
\end{figure}
 
The orbital analysis is based on three distinct models:
a circular single Keplerian, an eccentric single Keplerian,
and two circular Keplerians in a likely 2:1 mean motion resonance.
To compare the fit quality of the circular single-planet model
with that of the more complex models, we carried out 
a likelihood ratio test. 
Both, the eccentric single-planet, as well as the 
circular two-planet solution, provide a significantly better 
solution than the circular single-planet solution, which we
therefore rejected as a plausible explanation for the data.
The eccentric single-planet model and the two-planet model
are described by the same number of free parameters. As a matter
of fact, the eccentric single planet model yields a higher likelihood
and also a smaller jitter term
on the grounds of which it might be preferred. 
To further quantify this statement, 
we generated synthetic data sets based on the eccentric 
and the two-planet solution, and inspected the likelihood ratio 
distributions.
Our investigations show that none of the models can be rejected
on statistical grounds. As both
models are also physically plausible, we discuss their
implications below.

Based on the eccentric model,
GJ~4276~b has a minimum mass of $\sim$16.6\,M$_\oplus$,
an orbital period of 13.4 days, and is located closer than the 
inner edge of the habitable zone at 0.08\,au.
At this orbital distance the tidal circularization timescale for a gaseous planet
is more than 10\,Gyr, which is consistent with an eccentric orbit.
Analyzing the periodogram of the residual RVs, we find no immediate 
evidence for further planetary companions around GJ~4276.
We show in Fig.~\ref{figure:eccentricities} the eccentricity 
of the known exoplanets around M dwarfs as a function of orbital period.
There are 13 planetary systems with published eccentricities of $e \ge 0.3$.
With a relatively high eccentricity of $0.37\pm0.03$, 
GJ~4276~b would be among the most eccentric exoplanets around M dwarfs 
known to date and comparable with the recently published exoplanet 
GJ~96~b with $e=0.44^{+0.09}_{-0.11}$ \citep{Hobson2018}. 
However, while both planets have similar masses,
they differ significantly in the orbital period 
($m\sin i=19.66^{+2.42}_{-2.30}$\,M$_\oplus$ and $P=73.94^{+0.33}_{-0.38}$\,d
for GJ~96~b). 

A 2:1 mean motion resonance is found in many planetary systems such as
HD~82943, HD~128311, HD~73526, HD~90043, and HD~27894 
\citep{Mayor2004, Vogt2005, Tinney2006, Johnson2011, Trifonov2017}. 
So far, two systems with M~dwarf host stars and planets near the 2:1 resonance
are known, viz., GJ~876 \citep{Marcy2001, Rivera2010} and TRAPPIST-1 \citep{Gillon2017}.
Both of these systems harbor more than two known planets with orbital periods in
a 4:2:1 resonance chain for GJ~876 and 8:5:3:2:1 for TRAPPIST-1.
Given these examples, we consider a resonant two-planet system also
a plausible model for GJ~4276.
While we here focus on a strict 2:1 period ratio, we note that
a slightly larger period ratio around 2.2 is often realized
\citep{Steffen2015}.
Also, strictly zero eccentricity, as assumed in our modeling, discards the dynamical
mutual gravitational interaction between the planets, which is expected to lead to
small, periodically changing eccentricities in the system. However, we consider
this idealization of the two-planet model justified, to study the data set at hand.
According to our two-planet model with a period ratio of 2:1,
the planets GJ~4276~b and c have minimum masses 
of $m_{\mathrm{b}} \sin i = 15.6\,$M$_\oplus$ and 
$m_{\mathrm{c}} \sin i = 4.4\,$M$_\oplus$. 
The two planets
orbit their parent star at separations of $a_{\mathrm{b}}=0.08\,$au and 
$a_{\mathrm{c}}=0.05\,$au and have orbital periods of 
$P_{\mathrm{b}} = 2P_{\mathrm{c}} = 13.35$ days. Still, both
planets would be inward of the habitable zone. 

Based on our statistical analysis, we express some preference 
for the the single-planet eccentric solution. However, also the
two-planet mean motion resonance is physically plausible, albeit formally
less strongly backed by the data at hand.
Conclusive evidence for one or the other alternative requires the number of RV measurements 
to be increased with follow-up observations. 
Nevertheless, the GJ~4276 planetary system shows a special configuration,
making it a highly interesting object for follow-up studies.

\begin{acknowledgements}
CARMENES is an instrument for the Centro Astronómico Hispano-Alemán de Calar Alto (CAHA, Almería, Spain). CARMENES
is funded by the German Max-Planck-Gesellschaft (MPG), the Spanish Consejo Superior de Investigaciones Científicas (CSIC),
the European Union through FEDER/ERF FICTS-2011-02 funds, and the members of the CARMENES Consortium (Max-
Planck-Institut für Astronomie, Instituto de Astrofísica de Andalucía, Landessternwarte Königstuhl, Institut de Ciències de
l'Espai, Insitut für Astrophysik Göttingen, Universidad Complutense de Madrid, Thüringer Landessternwarte Tautenburg,
Instituto de Astrofísica de Canarias, Hamburger Sternwarte, Centro de Astrobiología and Centro Astron\'omico Hispano-
Alemáan), with additional contributions by the Spanish Ministry of Science through projects AYA2016-79425-C3-1/2/3-P,
AYA2015-69350-C3-2-P, ESP2017-87676-C05-02-R, ESP2014-54362P, and ESP2017-87143R,
the German Science Foundation through the
Major Research Instrumentation Programme and DFG Research Unit FOR2544 “Blue Planets around Red Stars”, the Klaus
Tschira Stiftung, the states of Baden-Württemberg and Niedersachsen, and by the Junta de Andalucía.
This work made use of observations collected at Sierra Nevada Observatory (SNO) supported by the 
Instituto de Astrof\'isica de Andaluc\'ia, CSIC, and from the LCOGT network. 

EN acknowledges support through DFG project CZ 222/1-1.

S.C. acknowledges support from DFG project SCH 1382/2-1 and SCHM 1032/66-1.

G.A-E research is funded via the STFC Consolidated Grants ST/P000592/1, and a Perren foundation grant.


This work was prepared using PyAstronomy. 

This research has also made use of the {\tt corner.py} package \citep{Foreman-Mackey2016}.

\end{acknowledgements}

\bibliographystyle{aa}
\bibliography{bibs}

\begin{appendix}
\clearpage

\section{Rotation period analysis}

\begin{figure*}
\begin{center}
\includegraphics[width=0.49\textwidth]{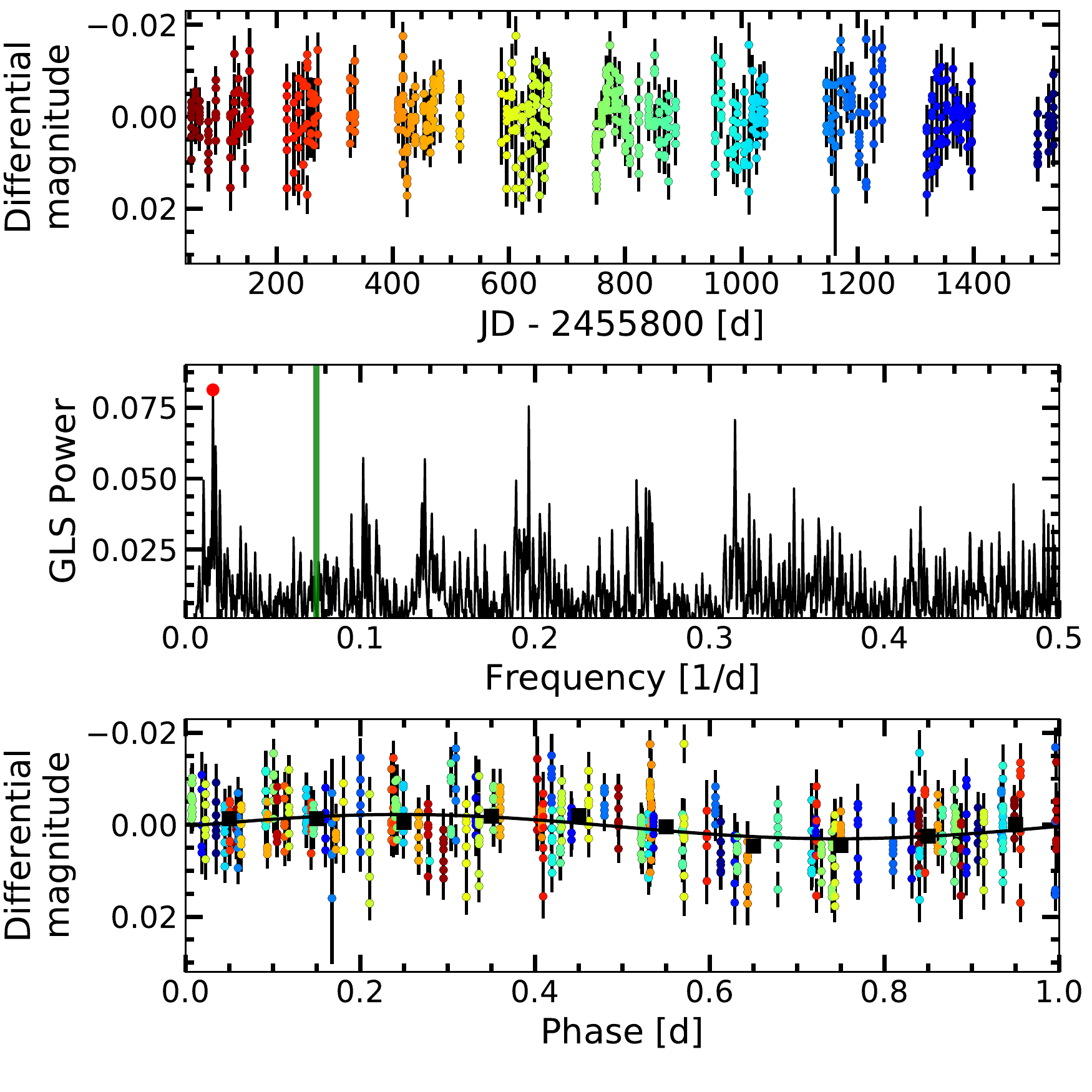}
\includegraphics[width=0.49\textwidth]{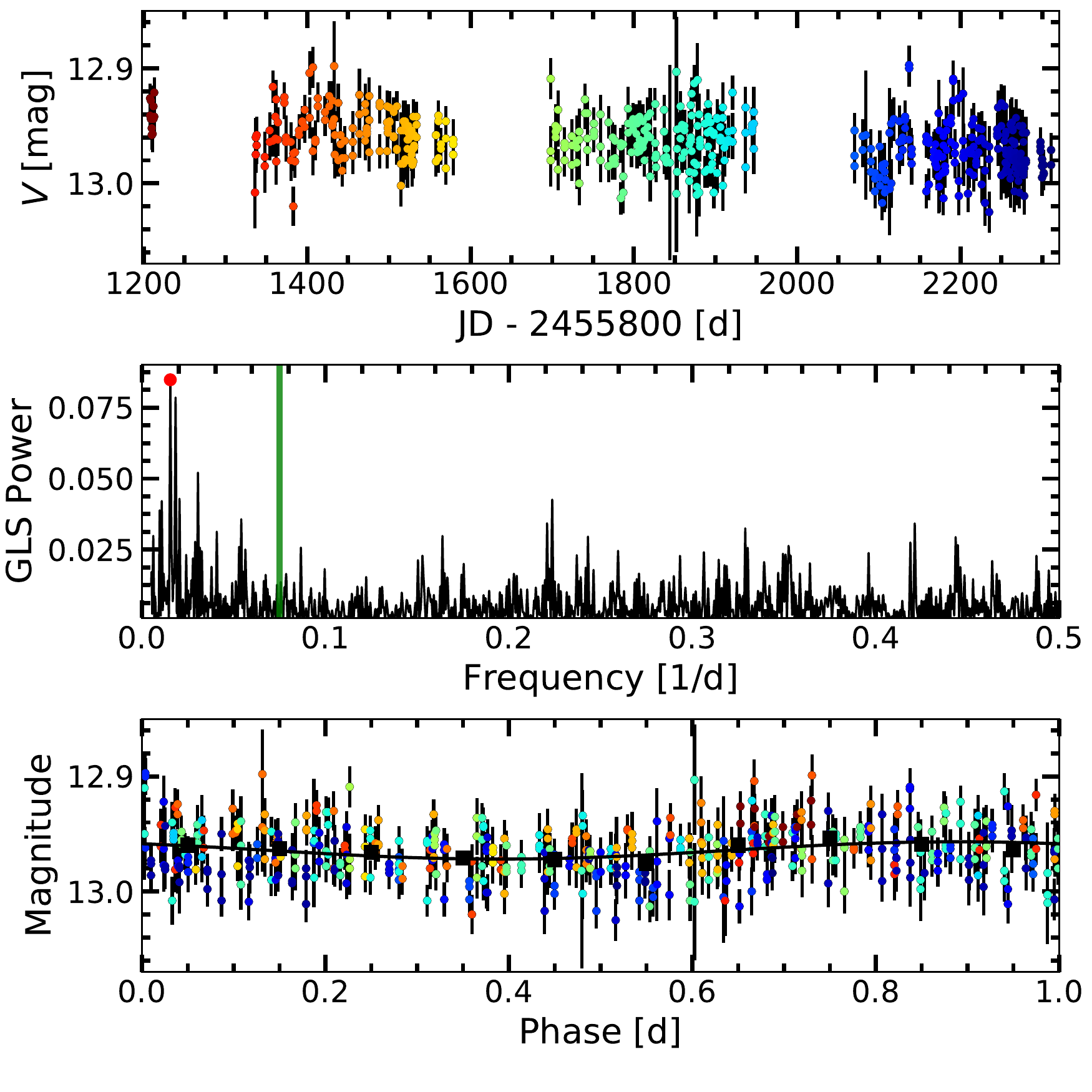}
\caption{Rotation period analysis using MEarth (\textit{left}) and ASAS-SN (\textit{right}) photometric data. 
\textit{Top:} RG715 broadband light curve (\textit{left}) and $V$ band light curve (\textit{right}). 
The color of the datapoints indicates the observation epoch.
\textit{Middle:} GLS periodograms. The vertical green lines represent the orbital period of the planet 
at 13.35 days and the red dots the peak with the highest power at 63.9 days (\textit{left}) and 
64.7 days (\textit{right}).
\textit{Bottom:} Phased light curves using the rotation period derived from the GLS.
The black curves show the best-fit sinusoidal models with an amplitude of 2.64\,mmag (\textit{left}) 
and 7.53\,mmag (\textit{right}).
The black squares indicate the mean magnitude in ten equidistant bins in phase.}
\label{figure:photometry_appendix}
\end{center}
\end{figure*}

\section{MCMC cornerplots}

\newpage
\begin{figure*}
\begin{center}
\includegraphics[width=\textwidth]{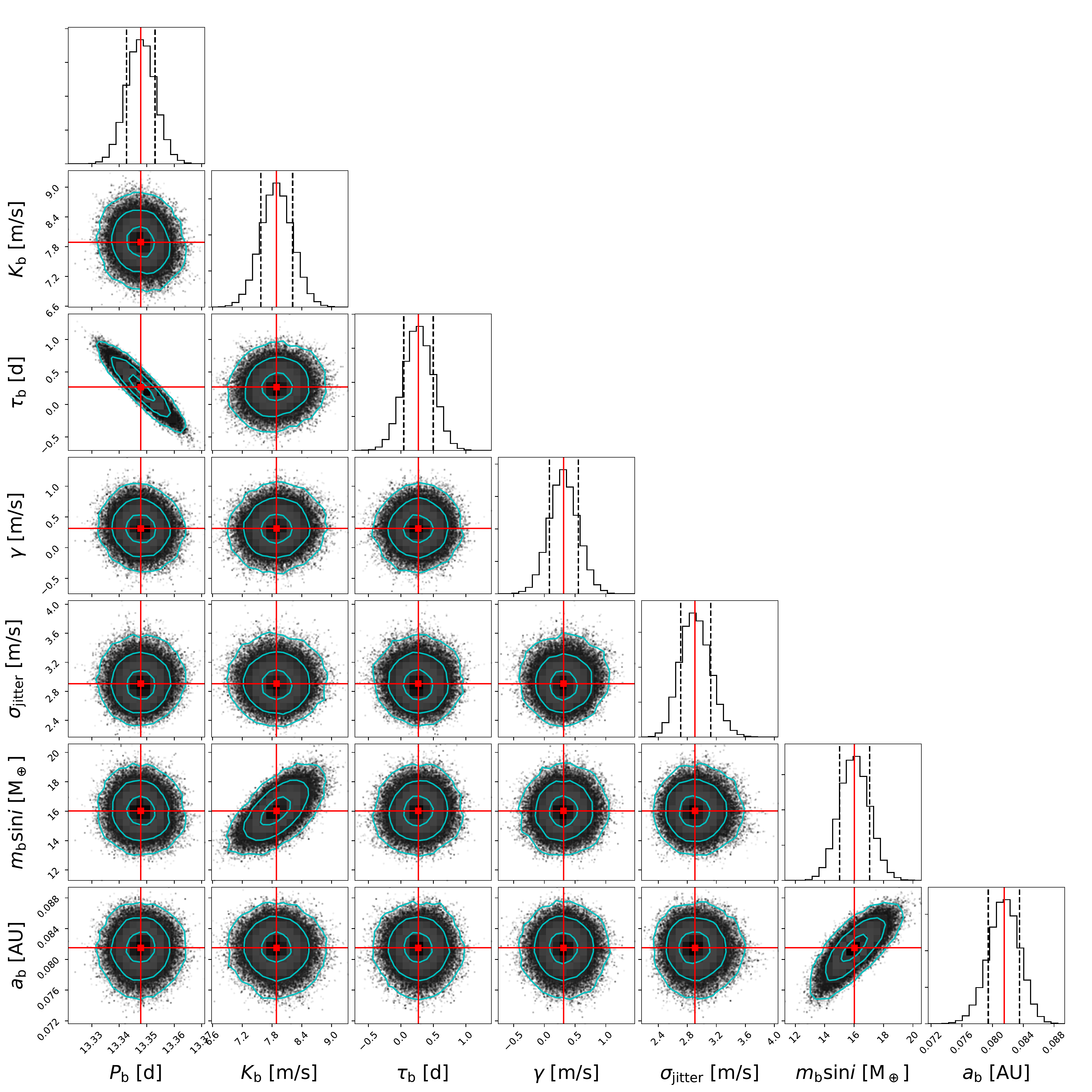}
\caption{
Two-dimensional projections of the posterior probability distributions of 
the circular single-planet Keplerian model. 
The contours represent the 1, 2, and 3$\sigma$ uncertainty levels. 
} 
\label{figure:corner1}
\end{center}
\end{figure*}

\begin{figure*}
\begin{center}
\includegraphics[width=\textwidth]{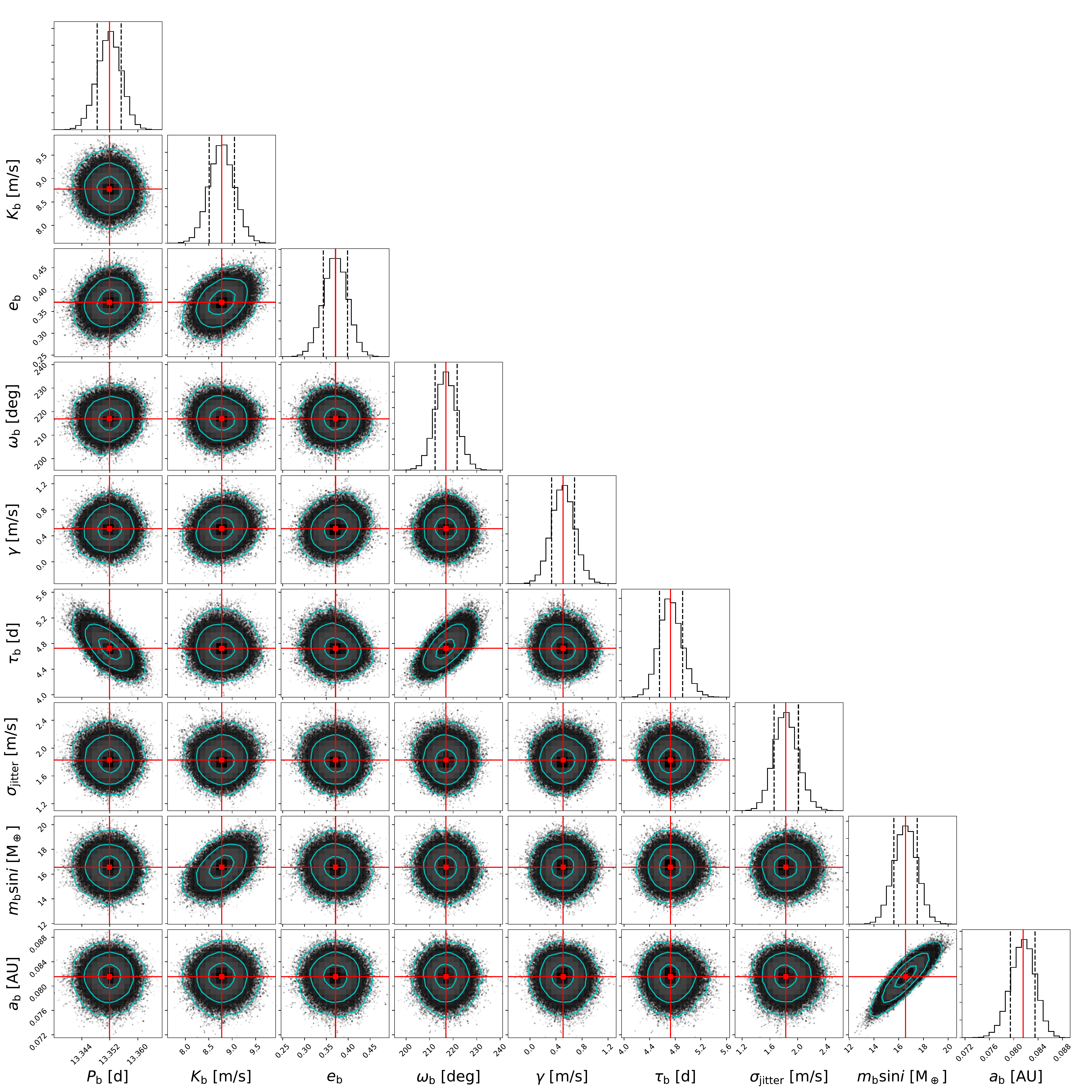}
\caption{
Same as Fig.~\ref{figure:corner1} but for 
the eccentric single-planet Keplerian model. 
} 
\label{figure:corner2}
\end{center}
\end{figure*}

\begin{figure*}
\begin{center}
\includegraphics[width=\textwidth]{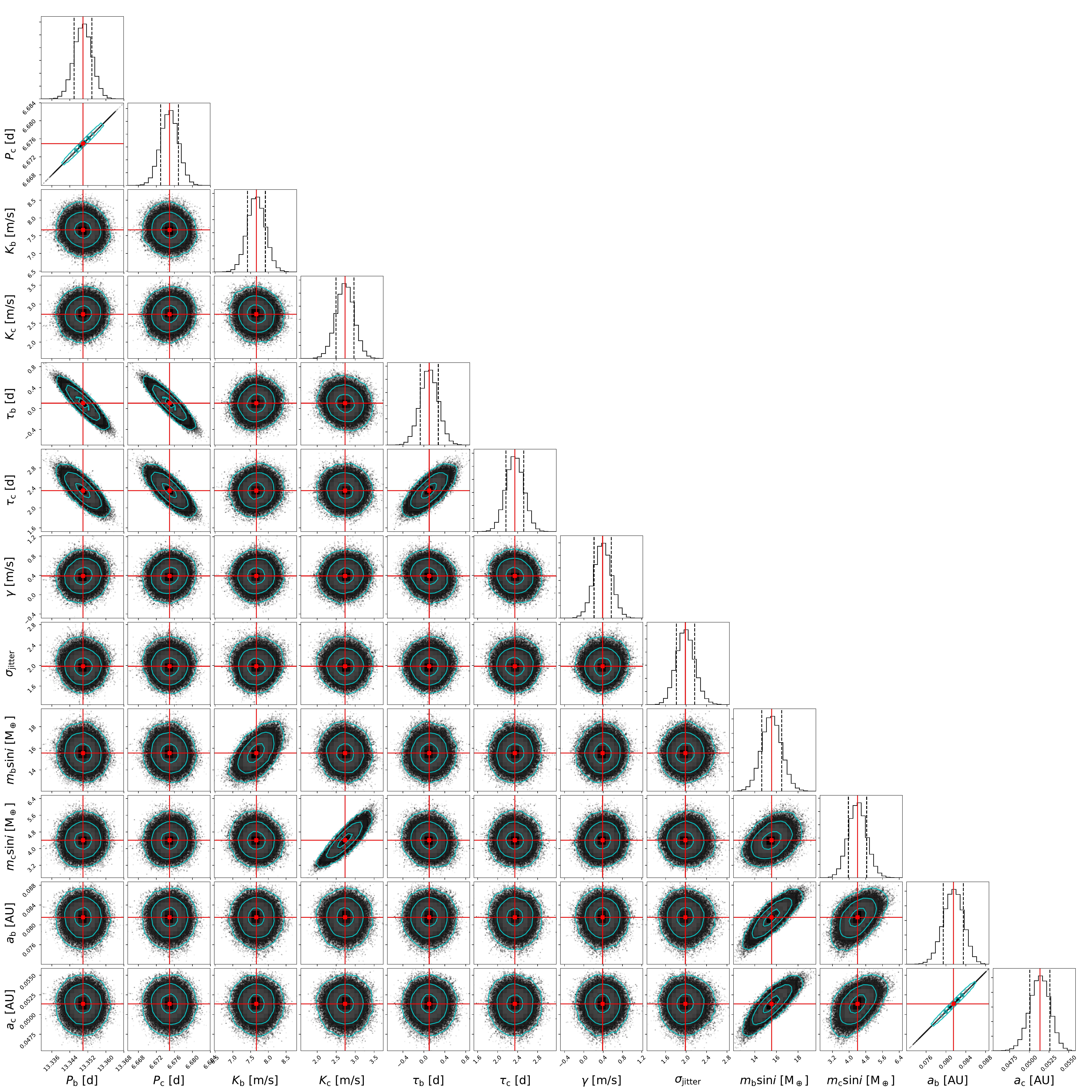}
\caption{
Same as Fig.~\ref{figure:corner1} but for the two-planet Keplerian model 
on circular orbits with a period ratio of 2:1. 
} 
\label{figure:corner3}
\end{center}
\end{figure*}

\section{Radial velocities of GJ~4276}
\label{appendix:rvs}

\begin{table}
\caption{Barycentric Julian date, radial velocities, and formal uncertainties for GJ~4276.}
\label{table:rv1}
\centering
\begin{tabular}{crc}
\hline\hline

BJD & RV [m\,s$^{-1}$] & $\sigma_{\mathrm{RV}}$ [m\,s$^{-1}$] \\
\hline
$ 2457572.656 $ & $ 3.60 $ & $ 2.03 $ \\
$ 2457592.602 $ & $ 3.37 $ & $ 2.36 $ \\
$ 2457603.619 $ & $ -8.50 $ & $ 3.11 $ \\
$ 2457610.546 $ & $ 3.53 $ & $ 1.72 $ \\
$ 2457612.517 $ & $ -3.72 $ & $ 1.91 $ \\
$ 2457618.462 $ & $ 0.21 $ & $ 2.04 $ \\
$ 2457619.542 $ & $ 5.18 $ & $ 1.69 $ \\
$ 2457620.491 $ & $ 4.62 $ & $ 1.70 $ \\
$ 2457621.515 $ & $ 4.97 $ & $ 1.69 $ \\
$ 2457622.477 $ & $ 7.59 $ & $ 1.62 $ \\
$ 2457628.501 $ & $ -8.30 $ & $ 1.44 $ \\
$ 2457629.505 $ & $ -14.72 $ & $ 1.98 $ \\
$ 2457643.487 $ & $ -6.80 $ & $ 1.47 $ \\
$ 2457937.636 $ & $ -7.59 $ & $ 1.61 $ \\
$ 2457948.647 $ & $ -7.85 $ & $ 1.64 $ \\
$ 2457954.573 $ & $ 7.69 $ & $ 1.66 $ \\
$ 2457960.578 $ & $ -2.08 $ & $ 1.39 $ \\
$ 2457961.548 $ & $ -5.11 $ & $ 1.27 $ \\
$ 2457969.522 $ & $ 3.71 $ & $ 1.38 $ \\
$ 2457970.466 $ & $ 6.85 $ & $ 1.47 $ \\
$ 2457975.539 $ & $ -7.87 $ & $ 1.96 $ \\
$ 2457976.584 $ & $ -12.65 $ & $ 1.77 $ \\
$ 2457977.512 $ & $ -9.26 $ & $ 1.54 $ \\
$ 2457982.533 $ & $ 2.94 $ & $ 1.51 $ \\
$ 2457986.547 $ & $ -3.51 $ & $ 1.57 $ \\
$ 2457999.432 $ & $ 0.29 $ & $ 1.85 $ \\
$ 2458000.497 $ & $ -4.17 $ & $ 1.66 $ \\
$ 2458001.470 $ & $ -4.53 $ & $ 1.36 $ \\
$ 2458002.655 $ & $ -10.19 $ & $ 1.46 $ \\
$ 2458008.516 $ & $ 1.83 $ & $ 1.54 $ \\
$ 2458009.622 $ & $ 1.33 $ & $ 1.48 $ \\
$ 2458026.548 $ & $ 3.43 $ & $ 1.97 $ \\
$ 2458029.417 $ & $ -4.00 $ & $ 1.77 $ \\
$ 2458032.565 $ & $ 6.92 $ & $ 2.10 $ \\
$ 2458033.375 $ & $ 7.82 $ & $ 1.87 $ \\
$ 2458034.555 $ & $ 9.38 $ & $ 2.20 $ \\
$ 2458035.498 $ & $ 6.22 $ & $ 2.51 $ \\
$ 2458047.456 $ & $ 2.70 $ & $ 1.80 $ \\
$ 2458048.458 $ & $ 4.55 $ & $ 1.55 $ \\
$ 2458050.361 $ & $ 2.97 $ & $ 1.96 $ \\
$ 2458052.457 $ & $ -1.19 $ & $ 1.44 $ \\
$ 2458053.418 $ & $ -1.22 $ & $ 1.75 $ \\
$ 2458055.536 $ & $ -6.62 $ & $ 1.47 $ \\
$ 2458059.511 $ & $ 10.06 $ & $ 1.75 $ \\
$ 2458079.404 $ & $ 0.41 $ & $ 1.75 $ \\
$ 2458084.341 $ & $ -10.03 $ & $ 1.55 $ \\
$ 2458092.477 $ & $ 5.58 $ & $ 1.36 $ \\
$ 2458093.333 $ & $ 3.55 $ & $ 1.19 $ \\
$ 2458110.362 $ & $ -9.05 $ & $ 1.76 $ \\
$ 2458118.407 $ & $ 3.71 $ & $ 1.66 $ \\
$ 2458120.361 $ & $ 0.99 $ & $ 2.67 $ \\
$ 2458121.269 $ & $ -2.72 $ & $ 1.48 $ \\
$ 2458122.275 $ & $ -8.60 $ & $ 1.36 $ \\
$ 2458123.268 $ & $ -7.80 $ & $ 1.38 $ \\
$ 2458132.329 $ & $ 7.91 $ & $ 3.89 $ \\
$ 2458134.272 $ & $ 0.05 $ & $ 2.15 $ \\
$ 2458135.303 $ & $ -5.76 $ & $ 1.84 $ \\
$ 2458136.306 $ & $ -8.59 $ & $ 1.41 $ \\
$ 2458138.319 $ & $ -1.71 $ & $ 2.32 $ \\
$ 2458139.319 $ & $ 2.36 $ & $ 1.93 $ \\

\hline
\end{tabular}
\end{table}

\begin{table}
\caption{Continued.}
\label{table:rv2}
\centering
\begin{tabular}{crc}
\hline\hline
BJD & RV [m\,s$^{-1}$] & $\sigma_{\mathrm{RV}}$ [m\,s$^{-1}$] \\
\hline
$ 2458140.321 $ & $ 8.12 $ & $ 1.83 $ \\
$ 2458141.387 $ & $ 7.82 $ & $ 1.63 $ \\
$ 2458143.277 $ & $ 4.20 $ & $ 1.62 $ \\
$ 2458144.278 $ & $ 6.19 $ & $ 2.82 $ \\
$ 2458149.292 $ & $ -9.06 $ & $ 1.91 $ \\
$ 2458159.288 $ & $ 8.75 $ & $ 2.62 $ \\
$ 2458249.639 $ & $ 8.22 $ & $ 1.82 $ \\
$ 2458263.662 $ & $ 10.49 $ & $ 2.94 $ \\
$ 2458270.656 $ & $ -9.49 $ & $ 2.39 $ \\
$ 2458284.631 $ & $ -6.51 $ & $ 1.69 $ \\
$ 2458290.631 $ & $ 6.73 $ & $ 1.94 $ \\
$ 2458291.634 $ & $ 4.76 $ & $ 1.75 $ \\
$ 2458292.638 $ & $ 1.13 $ & $ 1.56 $ \\
$ 2458293.645 $ & $ -0.85 $ & $ 1.48 $ \\
$ 2458296.629 $ & $ -14.35 $ & $ 1.83 $ \\
$ 2458297.620 $ & $ -8.54 $ & $ 1.51 $ \\
$ 2458300.615 $ & $ 4.96 $ & $ 2.00 $ \\
$ 2458303.611 $ & $ 10.32 $ & $ 2.84 $ \\
$ 2458304.630 $ & $ 4.17 $ & $ 2.94 $ \\
$ 2458305.624 $ & $ 4.64 $ & $ 1.79 $ \\
$ 2458306.622 $ & $ 4.94 $ & $ 1.86 $ \\
$ 2458309.589 $ & $ -9.87 $ & $ 1.90 $ \\
$ 2458313.607 $ & $ 3.33 $ & $ 1.39 $ \\
$ 2458315.596 $ & $ 7.10 $ & $ 2.95 $ \\
$ 2458316.596 $ & $ 5.01 $ & $ 1.68 $ \\
$ 2458317.624 $ & $ 8.02 $ & $ 2.07 $ \\
$ 2458318.617 $ & $ 6.41 $ & $ 2.03 $ \\
$ 2458324.627 $ & $ -7.49 $ & $ 1.60 $ \\
$ 2458326.621 $ & $ 4.03 $ & $ 1.40 $ \\
$ 2458327.603 $ & $ 9.50 $ & $ 1.41 $ \\
$ 2458330.590 $ & $ 8.47 $ & $ 1.29 $ \\
$ 2458332.595 $ & $ 4.96 $ & $ 1.27 $ \\
$ 2458336.603 $ & $ -9.74 $ & $ 1.53 $ \\
$ 2458337.671 $ & $ -8.30 $ & $ 1.49 $ \\
$ 2458338.509 $ & $ -5.47 $ & $ 1.64 $ \\
$ 2458339.580 $ & $ 4.10 $ & $ 1.64 $ \\
$ 2458340.620 $ & $ 6.97 $ & $ 1.35 $ \\
$ 2458343.563 $ & $ 4.52 $ & $ 1.52 $ \\
$ 2458345.561 $ & $ 1.04 $ & $ 1.43 $ \\
$ 2458346.542 $ & $ -1.60 $ & $ 1.53 $ \\
\hline
\end{tabular}
\end{table}

\end{appendix}
\end{document}